\documentclass[aps,prb,twocolumn,superscriptaddress,longbibliography]{revtex4-2}

\usepackage{nicefrac,xfrac}
\usepackage{graphicx}
\usepackage{dcolumn}
\usepackage{bm}
\usepackage{xcolor}
\usepackage{multirow}

\usepackage{makecell}
\usepackage{float}
\usepackage{amsmath}
\usepackage{color, colortbl}
\usepackage{soul}
\usepackage{subcaption}
\usepackage{parskip}
\usepackage{SIunits}
\usepackage[colorlinks,bookmarks=false,linkcolor=blue,urlcolor=blue]{hyperref}
\usepackage{enumitem}
\usepackage{dcolumn}

\begin{document}

\title{Dopant-induced modulation of ferroelectricity in perovskite nitride LaWN$_3$}

\author{Harshvardhan Singh Deora}
\email{dsharshvardh@iisc.ac.in}
\affiliation{Solid State and Structural Chemistry Unit, Indian Institute of Science, Bangalore 560012, India.}

\author{Awadhesh Narayan}
\email{awadhesh@iisc.ac.in}
\affiliation{Solid State and Structural Chemistry Unit, Indian Institute of Science, Bangalore 560012, India.}

\date{\today}

\begin{abstract}
Perovskite nitrides are starting to be explored for their promising properties distinct from their oxide counterparts. Here, through first-principles density functional computations, we study the intricate relationship between ferroelectricity and metallicity in the ferroelectric nitride LaWN\textsubscript{3}. We systematically assess the impact of electron and hole doping via the background charge method, revealing that both types of charge carriers diminish the propensity for ferroelectricity in LaWN\textsubscript{3}, although to remarkably different extents. Specifically, the introduction of electrons leads to centrosymmetry at a lower concentration value of nearly 0.2 electrons per formula unit. In contrast, the addition of holes does not result in centrosymmetry at reasonable doping values of up to 0.3 holes per formula unit. We present the underlying mechanisms behind these findings, noting that adding electrons fills the valence orbitals of both W and N atoms. This filling leads to a screening effect on the long-range repulsion between these atoms, thereby reducing the off-centering of W. On the other hand, hole doping does not significantly alter the W-N bonding nature, resulting in a more robust off-centering. Furthermore, we introduce explicit impurity atoms and analyze the influence of the different factors contributing to the change in polarization, namely, the size of the dopant, charge carriers introduced by the dopant, and changes in lattice constants. We discover that doped LaWN\textsubscript{3} can be a promising polar metal, especially with acceptor dopants. Our study comprehensively shows the interplay between polarity and metallicity in this prototypical perovskite nitride.
\end{abstract}

\maketitle


\section{Introduction}

Materials that exhibit polar symmetry and possess spontaneous electric polarization, which can be switched by an external electric field, are referred to as ferroelectric materials. The introduction of charge carriers in ferroelectric materials tends to screen the long-range interactions that promote ferroelectric structural distortions~\cite{lines2001principles}, thereby inhibiting polarization switching. This phenomenon often leads to the perception that ferroelectricity and metallicity are mutually exclusive properties. However, some time ago, it was theoretically demonstrated by Anderson and Blount that these properties can coexist in a material if the interaction between the polar distortions and the electrons is sufficiently weak~\cite{anderson1965symmetry}. Such materials are referred to as ``polar metals"~\cite{zhou2020review,bhowal2023polar,hickox2023polar}. Polar metals hold significant promise for a range of applications. Materials that maintain robust polarization in the presence of charge doping can serve as effective electrodes for stabilizing thin-film polarization~\cite{puggioni2018polar}. Some of them can show giant optical responses, which make them promising candidates for optoelectronic devices~\cite{kim2016polar}. Additionally, they are well-suited for designing materials with tunable metal–insulator transitions~\cite{puggioni2015design}. Polar metals may also be relevant to phenomena such as superconductivity and topological states~\cite{rischau2017ferroelectric,yu2018nonsymmorphic}. A well-known example is the perovskite oxide LiOsO\textsubscript{3}, which can exhibit a polar crystal class and possess a finite density of states at the Fermi level~\cite{shi2013ferroelectric}. Furthermore, polar metallicity in BaTiO\textsubscript{3} has been extensively studied~\cite{li2021strain,takahashi2017polar,ma2018strain,raghavan2016probing}, where the addition of charge carriers reduces the off-centering of Ti\textsuperscript{4+} due to the screening of the long-range coulombic attraction term~\cite{burdett1981use}, ultimately leading to the suppression of polarization~\cite{michel2021interplay}.

Oxide materials with the perovskite structure have been extensively studied for decades~\cite{cohen1992origin,pena2001chemical,ghosez2022modeling,yin2019oxide}. Although nitride perovskites have been theoretically predicted to be stable, their synthesis remains limited, and their properties require further exploration. Given that the standard oxidation state of nitrogen is $3-$, achieving this configuration would necessitate exceptionally high oxidation states for the cations, ranging from 5+ to 7+. While certain transition metals can sustain such elevated oxidation states, the lower electronegativity of nitrogen compared to oxygen suggests that achieving these high oxidation states is likely to be more challenging~\cite{sarmiento2015prediction}. Nitride perovskites may exhibit various emergent properties or concealed states due to the mixed covalent and ionic nature of the metal-nitrogen bonds, attributable to lower electronegativity of nitrogen relative to oxygen~\cite{disalvo1990solid}. In 2015, Sarmiento-Perez \textit{et al.} conducted a pioneering study to assess the thermodynamic feasibility of synthesizing nitride perovskites, specifically those with the general formula ABN\textsubscript{3}~\cite{sarmiento2015prediction}. Their research identified three nitride perovskites -- LaReN\textsubscript{3}, YReN\textsubscript{3}, and LaWN\textsubscript{3} -- as promising candidates for the perovskite structure. Notably, LaWN\textsubscript{3} was found to be a semiconductor with a significant spontaneous polarization. This finding was further experimentally validated by Talley \textit{et al.}, who synthesized LaWN\textsubscript{3} in the perovskite structure and confirmed its polar symmetry~\cite{talley2021synthesis}. Furthermore, several theoretical investigations have been conducted on LaWN\textsubscript{3}~\cite{nakaoka2024analyzing,an2024large,lai2024exploring,na2024improved,liu2020first,ren2023first}, with predictions extending to other nitride perovskites, including YMoN\textsubscript{3}, YWN\textsubscript{3}, ZrTaN\textsubscript{3}, and LaMoN\textsubscript{3}~\cite{grosso2023accessible}. In addition to these studies, some nitride perovskites have been experimentally reported, including TaThN\textsubscript{3}~\cite{brese1995synthesis}, LaReN\textsubscript{3}~\cite{kloss2021preparation}, and GdWN\textsubscript{3}~\cite{smaha2024gdwn}.

In this paper, we present a detailed investigation of the effects of doping on the prototypical nitride perovskite LaWN\textsubscript{3}, focusing on the structural, electronic, and polar properties. We begin by outlining the computational parameters and approximations employed in our calculations in Sec.~\ref{sec:Computational details}. Sec.~\ref{sec:Results and discussions} is dedicated to the discussion and explanation of our results. We start by briefly discussing the preliminaries in Sec.~\ref{sec:Preliminaries}, where we revisit the symmetry-adapted structural phase transitions in LaWN\textsubscript{3}. Following this analysis of the symmetry modes, we address the central question of how charge doping influences the polar distortions in LaWN\textsubscript{3}, which serves as the primary motivation for this study. We specifically examine how the off-centering of tungsten (W) atoms is affected by the introduction of charge carriers. Our approach first considers electron and hole doping via the background charge method, as detailed in Sec.~\ref{sec:Charge doping}, followed by the study of explicit impurity atoms or dopants in Sec.~\ref{sec:Impurity atoms in supercells}. Finally, we summarize our findings in Sec.~\ref{sec:Conclusions}.

\section{Computational details}
\label{sec:Computational details}

\subsection{Density functional theory calculations}

First-principles density functional theory (DFT) calculations are performed using the Quantum Espresso package~\cite{giannozzi2009quantum,giannozzi2017advanced}, employing a plane-wave basis set. Ultrasoft pseudopotentials are utilized for La, W, N, and other dopant elements to describe the core electrons. The local density approximation (LDA) is used to account for the exchange-correlation effects~\cite{ceperley1980ground}. A plane-wave cutoff of 680 eV is utilized. Monkhorst-Pack $k$-point meshes~\cite{monkhorst1976special} of dimensions $8 \times 8 \times 8$ and $5 \times 5 \times 5$ are employed for the unit cell, and $2 \times 2 \times 2$ for the supercell calculations, respectively. The densities of states are computed using a $16 \times 16 \times 16$ $k$-point grid. Full structural relaxation is performed until the Hellmann-Feynman forces on each atom converge to below $5 \times 10^{-5}$ eV/\AA.  In rhombohedral notation, the lattice parameters for the ground state structure (R3c) are calculated as \(a = b = c = 5.58 \ \text{\AA}\), with an angle of 60.48$^{\circ}$, consistent with previous theoretical and experimental studies~\cite{fang2017lattice,talley2021synthesis}. 

Phonon calculations are conducted using $2 \times 2 \times 2$ supercells, and phonon band structures at arbitrary $q$-vectors are determined using the Phonopy code~\cite{togo2015first}.

\subsection{Charge doping}

Electrons were incorporated into the high-symmetry (R$\overline{3}$c) and low-symmetry (R3c) structures of LaWN\textsubscript{3} using the ``tot\_charge" flag in Quantum Espresso~\cite{giannozzi2009quantum,giannozzi2017advanced}. This introduces a compensating background charge. Subsequently, atomic positions were relaxed while maintaining the lattice constants fixed to their values from the undoped structures, as the relaxation of lattice parameters is not well-defined in this process~\cite{bruneval2015pressure,mascello2020theoretical}.

\subsection{Polarization and polar distortions}

We are interested in doped systems, which are metallic. In such cases, the Berry phase polarization is undefined~\cite{king1993theory,resta1993macroscopic}. To get an estimate of the ``polarization", we calculate the size of the polar distortion or polarization, \( P \), by summing the displacements, \( d_i \), of the ions from their reference positions in the high-symmetry structure, each multiplied by their Born effective charges (BECs), \( Z^*_i \)~\cite{spaldin2012beginner},

\begin{equation}\label{eq:BEC_P}
P = \frac{e}{\Omega}\sum_iZ_i^*d_i,
\end{equation}

where \( e \) represents the elementary electronic charge and \( \Omega \) denotes the unit cell volume. We utilize the BEC values for the ferroelectric (FE) structure provided in Reference~\cite{an2024large}: \( Z_{\text{La}}^* = +4.44 \), \( Z_{\text{W}}^* = +8.69 \), and \( Z_{\text{N}}^* = -4.38 \). Our computed ``polarization" values reflect the degree of ferroelectric-like structural distortion present in the system. We note that, in this article, the term ``polarization" refers to this effective polarization or non-centrosymmetric structural distortion, without implying switchability of the polarization.

\subsection{Orbital populations}

Crystal orbital Hamilton populations (COHPs) and integrated crystal orbital Hamilton populations (ICOHPs) were computed using the LOBSTER code~\cite{dronskowski1993crystal,maintz2016lobster,nelson2020lobster} to investigate the variations in bond strength between W and N upon charge doping. The basis sets employed for these calculations included W ($5s$, $6s$, $5p$, $6p$, $5d$) and N ($2s$, $2p$). Projector-augmented wave (PAW) potentials were utilized as LOBSTER supports only PAW pseudopotentials. The -ICOHP plots denote the changes in bonding strength as a function of doping. All calculations were performed on a $16 \times 16 \times 16$ $k$-point grid.

\section{Results and discussions}
\label{sec:Results and discussions}

\subsection{Preliminaries}
\label{sec:Preliminaries}

\begin{figure}
  \centering
  \includegraphics[width=1\linewidth]{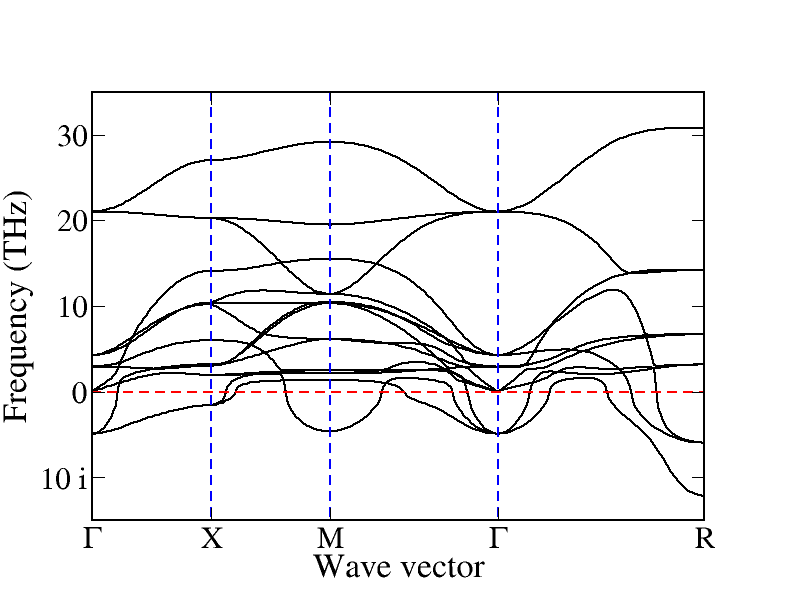}
    \caption{Phonon dispersion of the cubic structure (Pm$\overline{3}$m) of LaWN\textsubscript{3} along the high-symmetry path $\Gamma \rightarrow$ X $\rightarrow$ M $\rightarrow \Gamma \rightarrow$ R. It exhibits imaginary transverse optical modes at $\Gamma$ and antiferrodistortive modes at M and R, suggesting ferrodistortive distortions and octahedral rotations, respectively.}
  \label{fig:Phonon}
\end{figure}

The ground state of LaWN\textsubscript{3} is predicted to exhibit a structure with R3c symmetry, which can be derived from the cubic structure (Pm$\overline{3}$m) through two distinct phase transition pathways: Pm$\overline{3}$m $\rightarrow$ R$\overline{3}$c $\rightarrow$ R3c and Pm$\overline{3}$m $\rightarrow$ R3m $\rightarrow$ R3c~\cite{fang2017lattice}. The structural distortions associated with each phase transition can be elucidated through symmetry-adapted modes. Notably, despite the Goldschmidt tolerance factor being less than 1 ($t=0.969$)~\cite{liu2020first}, the observed polar instability in this material is primarily driven by the displacement of the B-site atoms, rather than the A-site atoms. This behavior contrasts with well-known ferroelectric materials, such as LiNbO\textsubscript{3}-type semiconductors (e.g., ZnSnO\textsubscript{3})~\cite{inaguma2008polar} and ferroelectric metals, such as LiOsO\textsubscript{3}~\cite{xiang2014origin,liu2015metallic,narayan2019effect}. The phonon dispersion of the cubic phase (Pm$\overline{3}$m) of LaWN\textsubscript{3}, shown in Fig.~\ref{fig:Phonon}, reveals that the cubic phase is dynamically unstable due to the imaginary frequency modes around the $\Gamma$, M, and R points. Strongly unstable antiferrodistortive modes at the R and M points are associated with the rotation of WN\textsubscript{6} octahedra. Additionally, three degenerate transverse optical modes exhibit large imaginary frequencies at the $\Gamma$ point, suggesting a potential ferroelectric transition. The eigenvectors of these $\Gamma$ point modes correspond to a displacement pattern in which N anions move in the opposite direction to the cations.

\begin{figure}
  \centering
  \begin{subfigure}{0.22\textwidth}
    \centering
     \includegraphics[width=\linewidth]{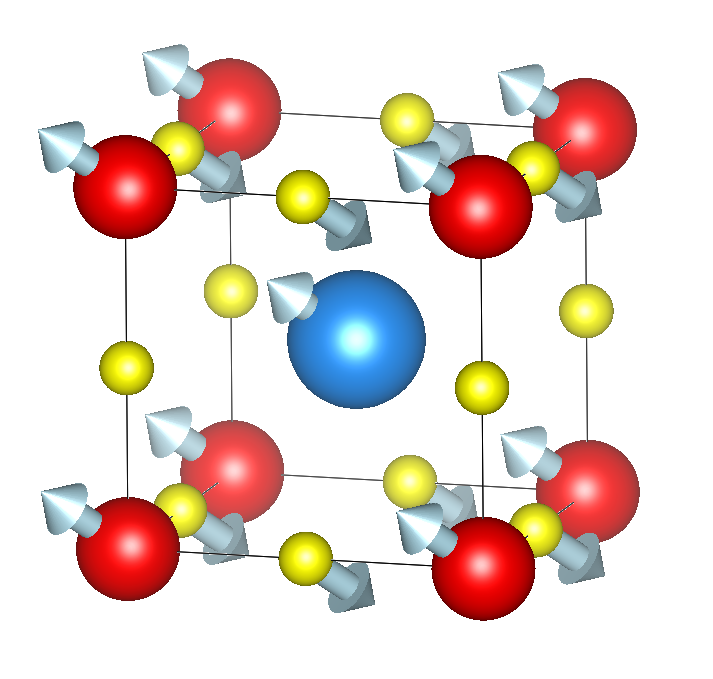}
    \caption{Pm$\overline{3}$m}
    \label{fig:cub_arrows}
  \end{subfigure}
  \hfill
  \begin{subfigure}{0.22\textwidth}
    \centering
     \includegraphics[width=\linewidth]{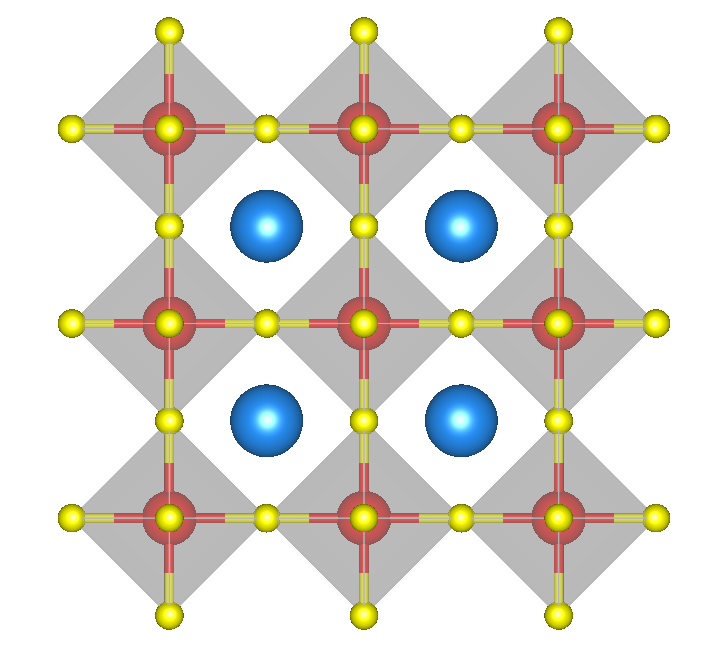}
    \caption{Pm$\overline{3}$m}
    \label{fig:cub}
  \end{subfigure}
  \hfill
 \begin{subfigure}{0.23\textwidth}
    \centering
     \includegraphics[width=\linewidth]{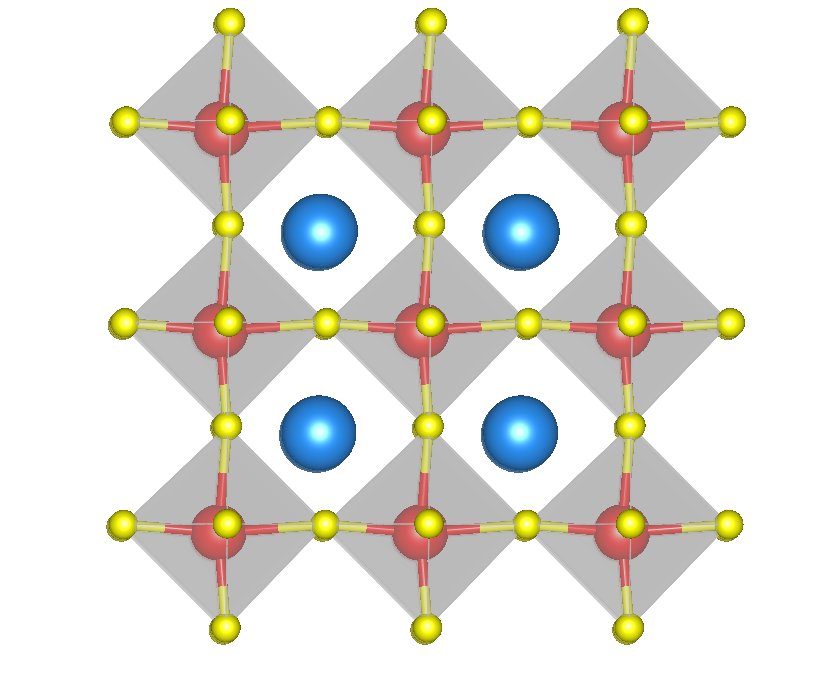}
    \caption{R3m}
    \label{fig:R3m}
  \end{subfigure}
  \hfill
 \begin{subfigure}{0.24\textwidth}
    \centering
     \includegraphics[width=\linewidth]{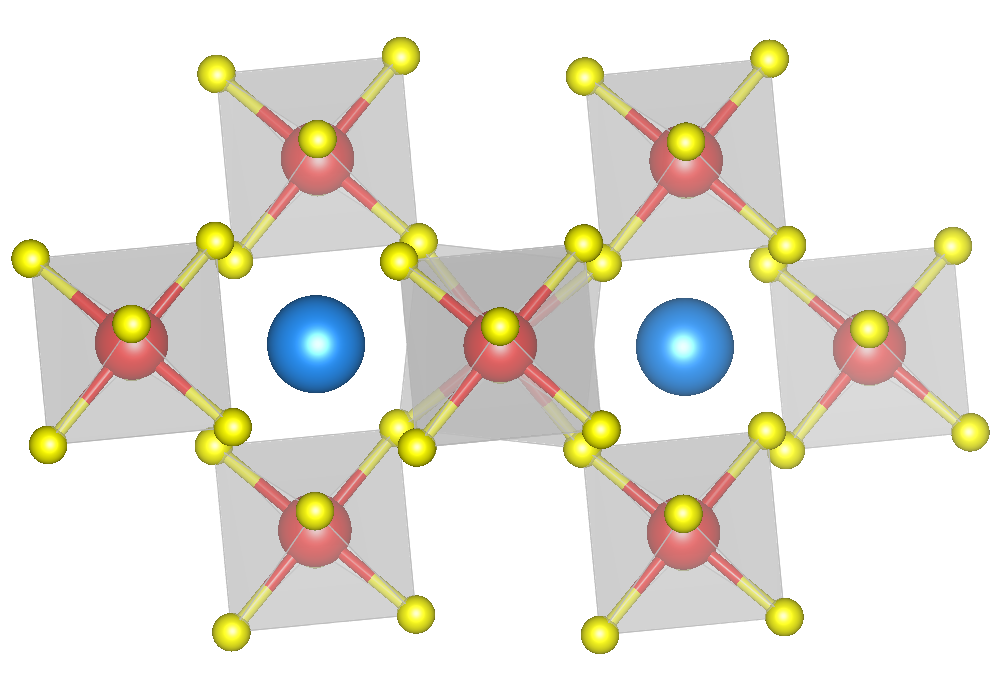}
    \caption{R$\overline{3}$c}
    \label{fig:R3barc}
  \end{subfigure}
  \hfill
 \begin{subfigure}{0.48\textwidth}
    \centering
     \includegraphics[width=\linewidth]{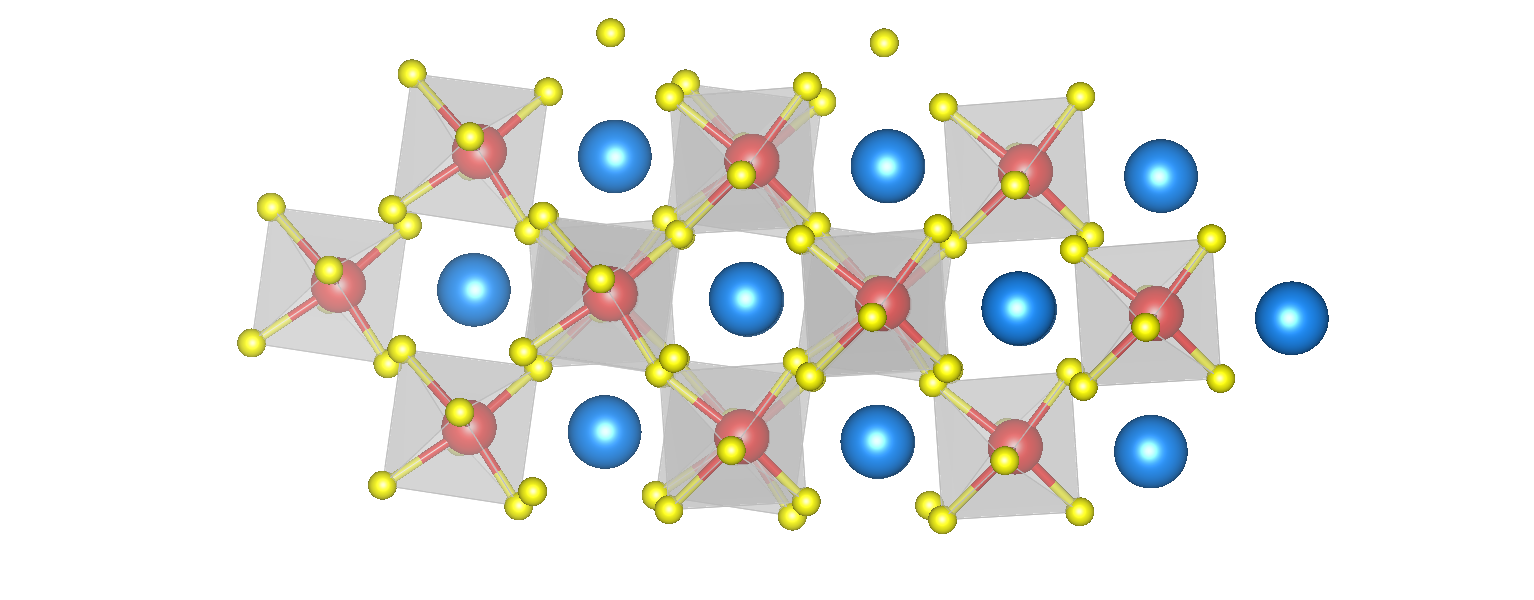}
    \caption{R3c}
    \label{fig:R3c}
  \end{subfigure}
  \hfill
  \caption{Panels (a) and (b) present the cubic structure of LaWN\textsubscript{3}, with La, W, and N depicted in blue, red, and yellow, respectively. In panel (a), the arrows on the atoms illustrate the polar distortion of W and La in the opposite direction to N along the pseudocubic $\langle 111 \rangle$ direction, which stabilize the imaginary modes at the $\Gamma$ point. Note that these arrows do not represent the quantitative magnitude of the distortions; the magnitude of distortion of La is much less compared to the other atoms due to its heavier mass. (c) Polar R3m structure of LaWN\textsubscript{3} that results from the polar distortion depicted in (a).  (d) Structure characterized by $a^{-} a^{-} a^{-}$ tilting of WN\textsubscript{6} octahedra, leading to the non-polar R$\overline{3}$c symmetry. (e) Ground state structure of LaWN\textsubscript{3}, which exhibits polar R3c symmetry involving both polar distortion and octahedra tilting.}
  \label{fig:Structures}
\end{figure}

To gain a deeper understanding of the unstable modes, we initially examined the effects of freezing ferrodistortive modes at the $\Gamma$ point, both individually and in combination. This analysis indicates that polar distortions of W and N, oriented oppositely along the pseudocubic $\langle 111 \rangle$ directions (as illustrated by arrows in Fig.~\ref{fig:cub_arrows}), minimize the system's energy. The resulting distorted structure exhibits R3m polar symmetry due to the breaking of inversion symmetry, as shown in Fig.~\ref{fig:R3m}. Similarly, when freezing antiferrodistortive modes, we find that the WN\textsubscript{6} octahedra tilt about their triad axis. This distortion can be described, in the Glazer notation~\cite{glazer1972classification}, as $a^{-} a^{-} a^{-}$ tilting, where adjacent octahedra rotate in opposite directions. The structure of LaWN\textsubscript{3} with the octahedra tilting assumes R$\overline{3}$c symmetry, which is non-polar due to the presence of an inversion center. Finally, by sequentially freezing both ferrodistortive and antiferrodistortive modes, we obtain the ground state with R3c polar symmetry, which exhibits no phonon instabilities~\cite{fang2017lattice}.

\subsection{Charge doping}
\label{sec:Charge doping}

Next, we begin our study of the effects of charge doping on this nitride perovskite system. To begin our investigation, we introduced charges via the background-charge method. This approach enables us to isolate and understand the influence of charge carriers independently from other variables, such as ion size or changes in valence orbitals that may accompany the introduction of explicit atoms.

\begin{figure}
  \centering
  \includegraphics[width=1\linewidth]{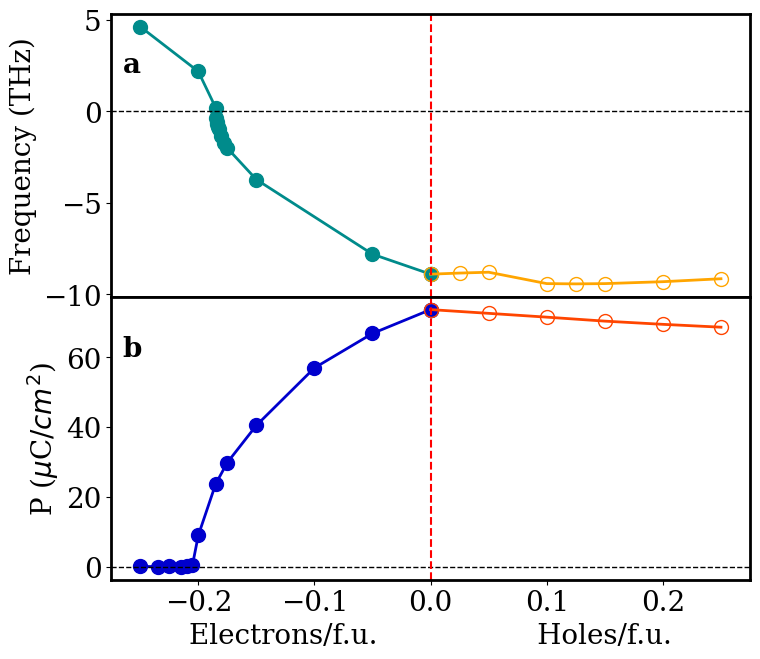}
  \caption{(a) The change in the frequency of the ferrodistortive mode at the $\Gamma$ point in the R$\overline{3}$c structure with charge doping. It takes approximately 0.2 electrons per formula unit (f.u.) for the least stable mode at the $\Gamma$ point to get stabilized. However, the addition of holes does not significantly affect the frequency of this transverse optical mode. Note that negative values of frequency represent the imaginary frequency of the ferrodistortive mode. (b) The change in polarization of the R3c structure with charge doping. These results are consistent with (a), as addition of electrons suppresses the polarization, while holes have a negligible effect. The vertical dashed line denotes the charge neutrality condition.}
  \label{fig:Back_ground_charge}
\end{figure}

\begin{figure}
  \centering

  \begin{subfigure}{0.23\textwidth}
    \centering
    \includegraphics[width=\linewidth]{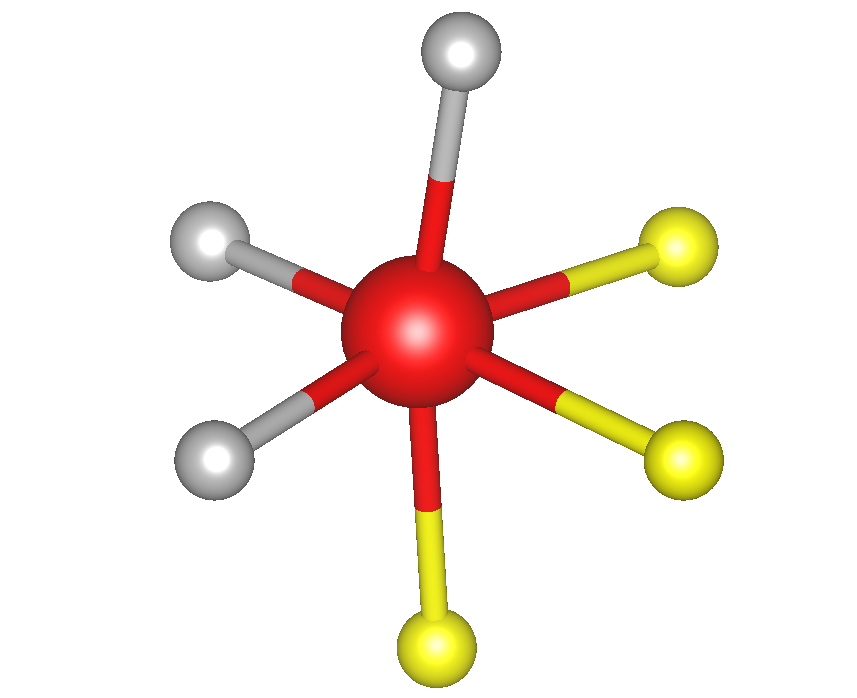}
    \caption{Undoped}
    \label{fig:dist}
  \end{subfigure}
  \hfill
  \begin{subfigure}{0.23\textwidth}
    \centering
    \includegraphics[width=\linewidth]{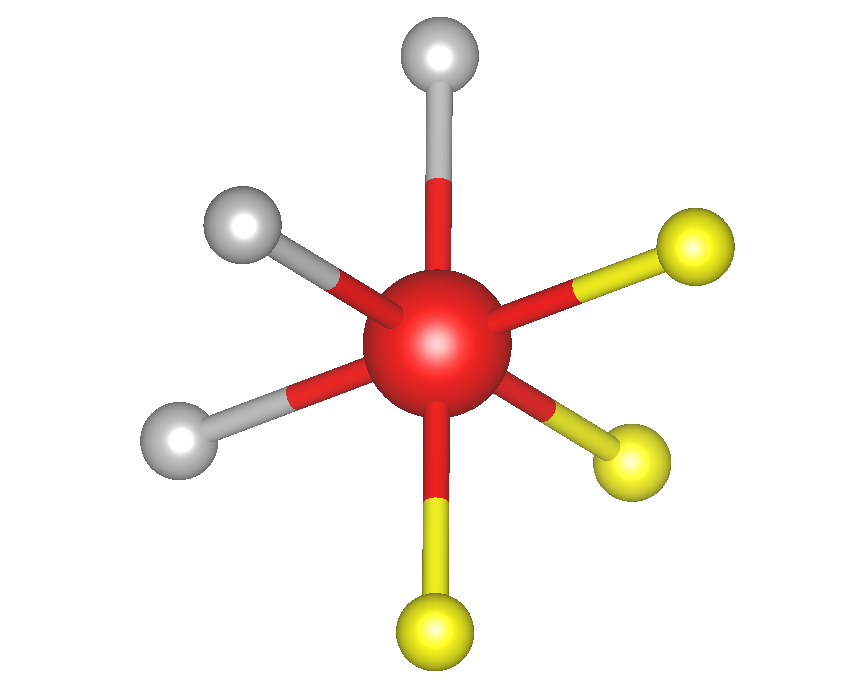}
    \caption{Doped}
    \label{fig:undist}
  \end{subfigure}
  \hfill
  \caption{The WN\textsubscript{6} octahedron, where W is represented by red color. Three cis axial-up nitrogen atoms are shown in grey color, and the three cis axial-down nitrogen atoms are shown in yellow color. (a) As discussed in Fig.~\ref{fig:Structures}, W and N are displaced opposite to each other along the $\langle 111 \rangle$ direction. This results in W being off-centered towards all three grey-colored nitrogen atoms, with equal bond lengths to each of them. The W-N (axial up) bond length is shorter than the W-N (axial down) bond length, and all three cis nitrogen atoms in yellow also have the same bond distance with W. For clarity, we use the term `axial up N' for the grey-colored nitrogen atoms and `axial down N' for the yellow-colored nitrogen atoms in the discussion. (b) When charge carriers are introduced, W exhibits a tendency to regain centrosymmetry. After a critical concentration of added charge carriers, the off-centering of W is entirely suppressed, resulting in equal bond distances between W and all surrounding nitrogen atoms.}

\end{figure}

\begin{figure*}[t]
  \centering

  \begin{subfigure}{0.339\textwidth}
    \centering
    \includegraphics[width=\linewidth]{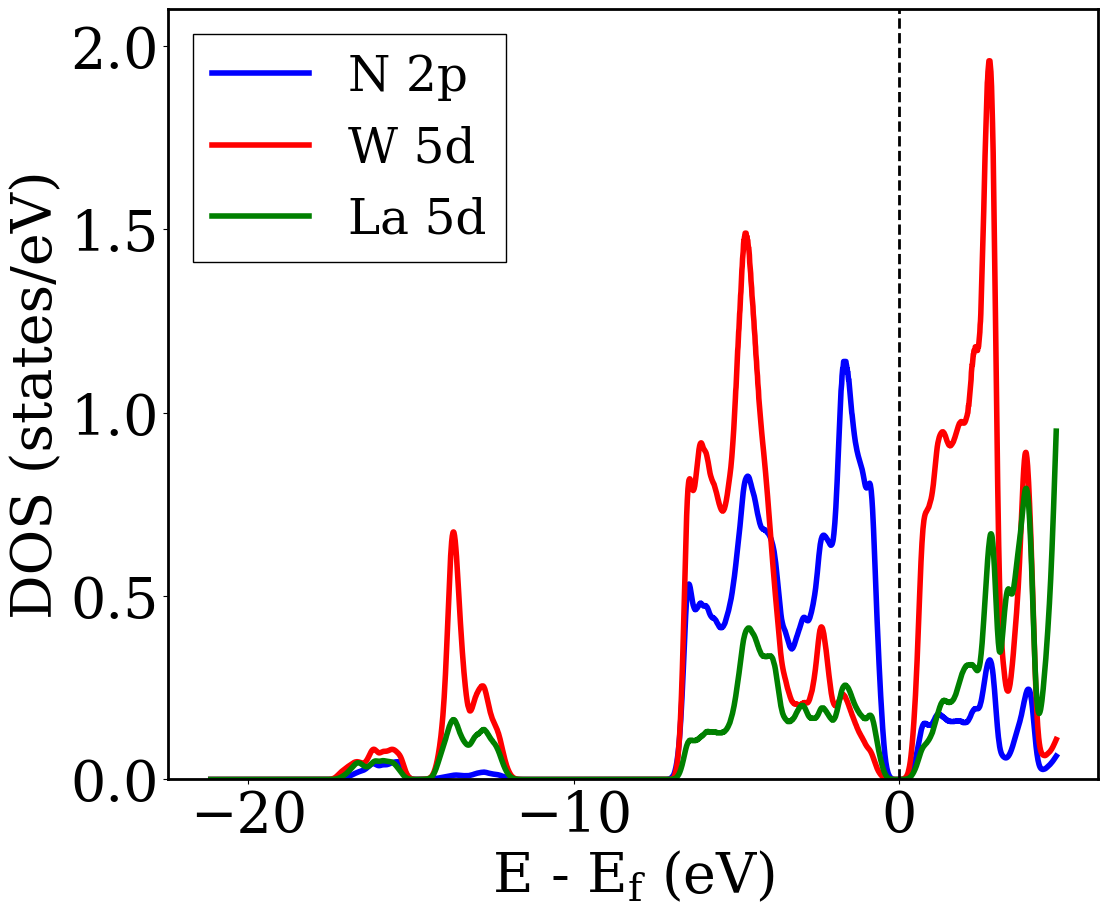}
    \caption{}
    \label{fig:DOS}
  \end{subfigure}
  \hfill
  \begin{subfigure}{0.32\textwidth}
    \centering
    \includegraphics[width=\linewidth]{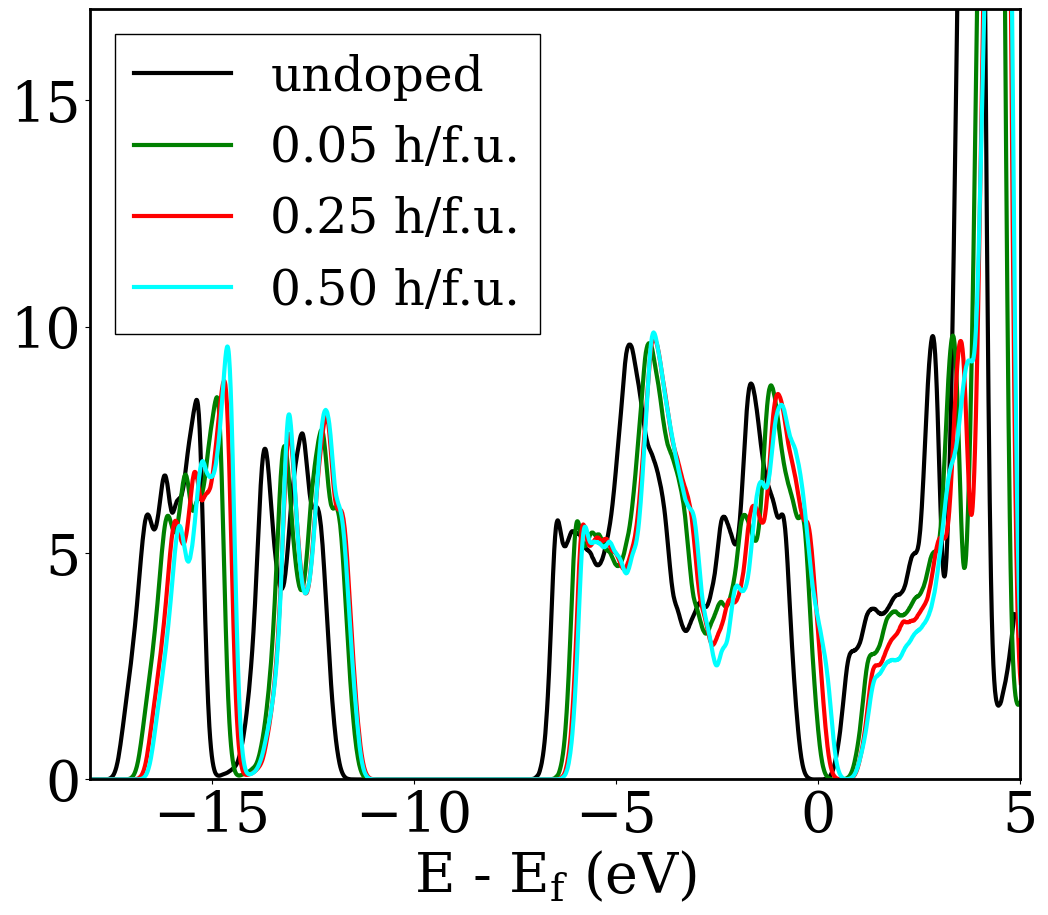}
    \caption{}
    \label{fig:DOS_h}
  \end{subfigure}
  \hfill
  \begin{subfigure}{0.32\textwidth}
    \centering
    \includegraphics[width=\linewidth]{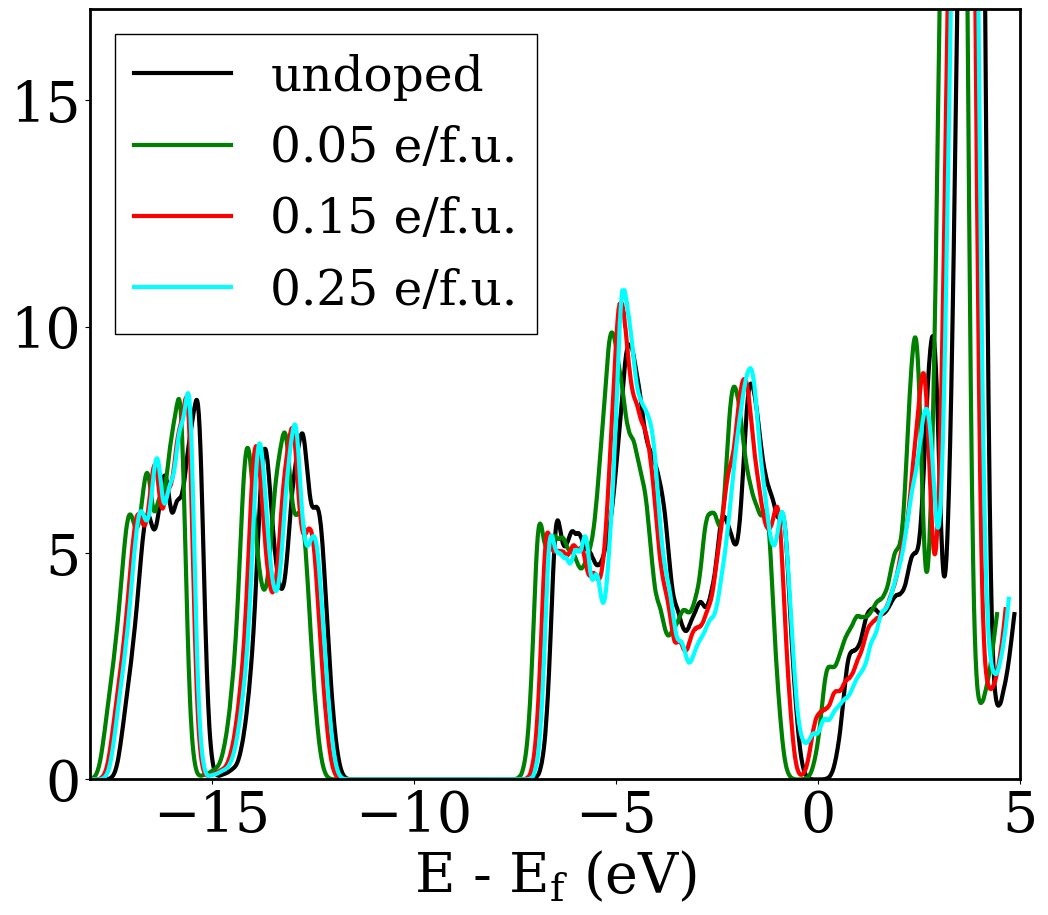}
    \caption{}
    \label{fig:DOS_e}
  \end{subfigure}
  \begin{subfigure}{0.345\textwidth}
    \centering
    \includegraphics[width=\linewidth]{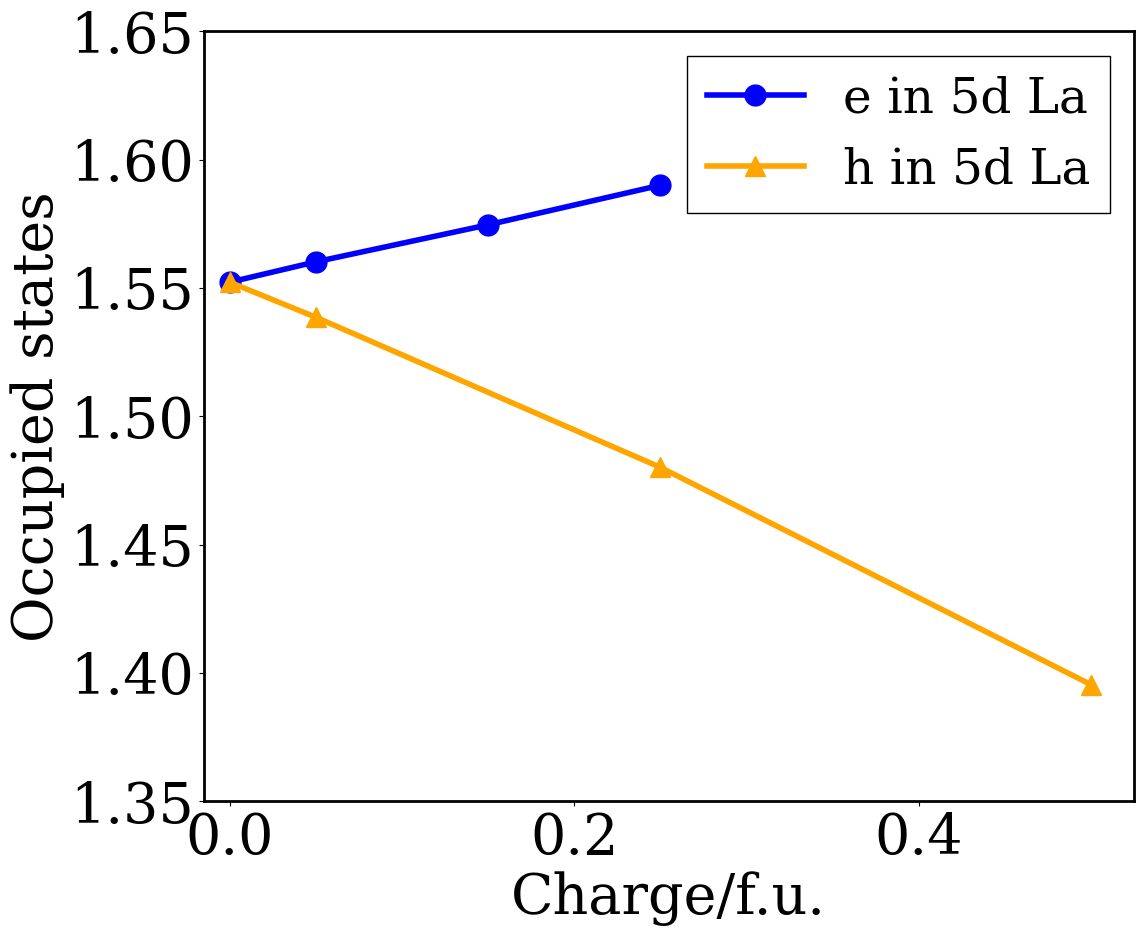}
    \caption{}
    \label{fig:occ_La}
  \end{subfigure}
  \begin{subfigure}{0.315\textwidth}
    \centering
    \includegraphics[width=\linewidth]{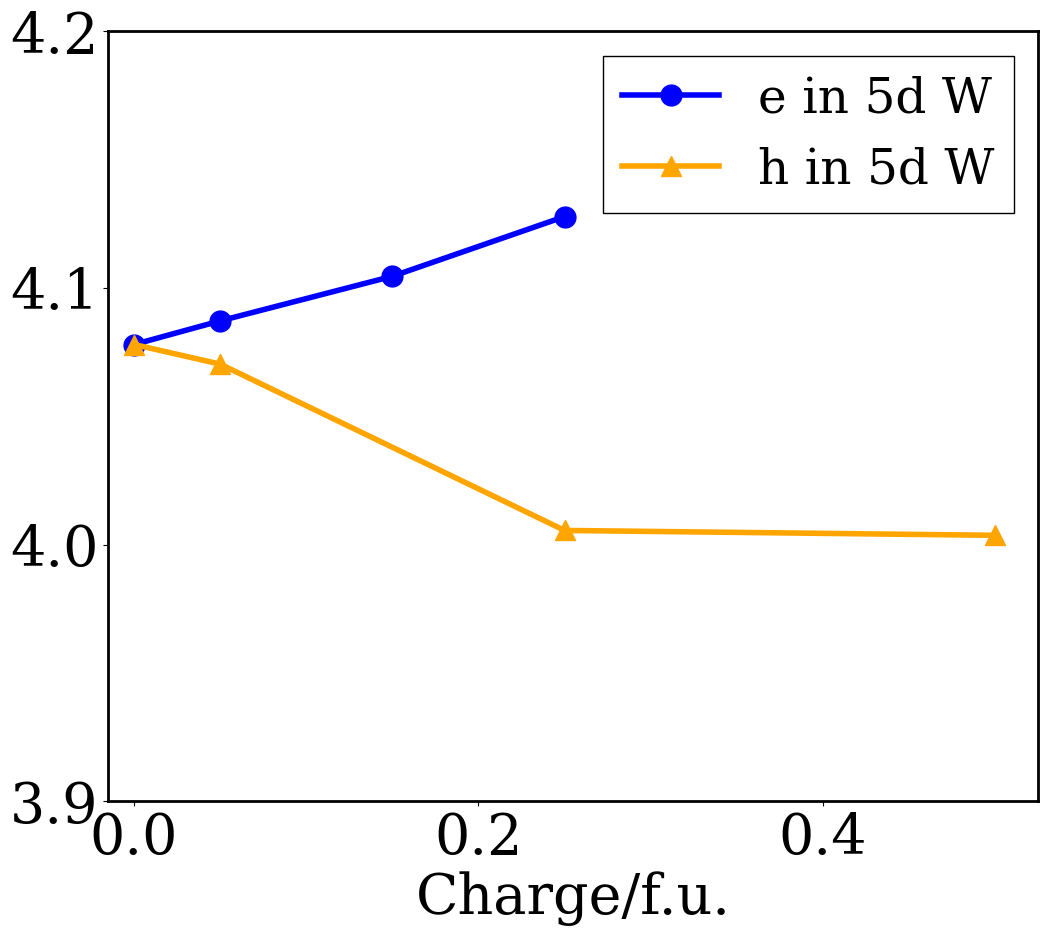}
    \caption{}
    \label{fig:occ_W}
  \end{subfigure}
  \begin{subfigure}{0.325\textwidth}
    \centering
    \includegraphics[width=\linewidth]{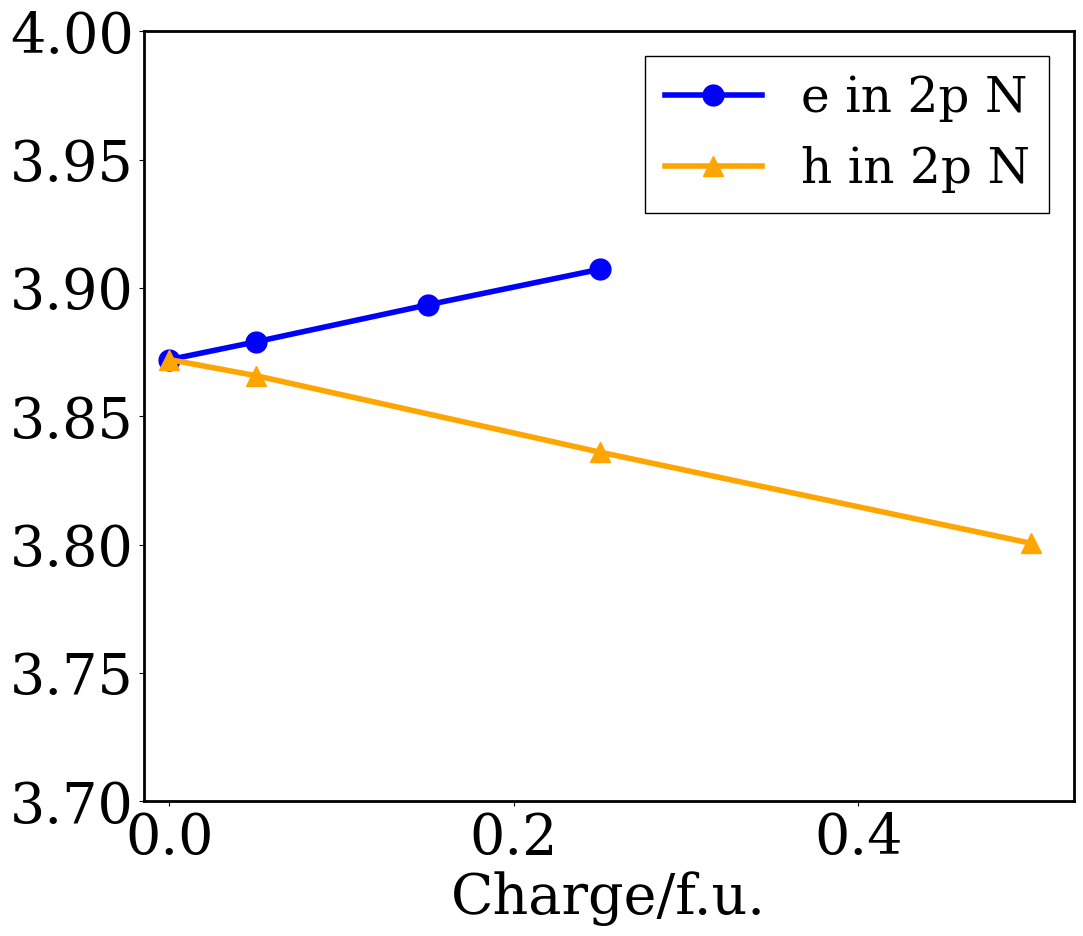}
    \caption{}
    \label{fig:occ_N}
  \end{subfigure}  
  \caption{(a) Density of states (DOS) of the $5d$ orbitals of La, $5d$ orbitals of W, and $2p$ orbitals of N. While all orbitals exhibit significant contributions to the states around the Fermi level, the valence band maximum is predominantly characterized by the $2p$ orbitals of N, whereas the conduction band minimum is mainly comprised of the $5d$ orbitals of W. (b) Variation in total DOS with hole doping, demonstrating a shift of the Fermi level into the valence band. (c) Variation in total DOS with electron doping, showing an initial rigid shift of the Fermi level at low electron concentrations, followed by a diminishing band gap beyond approximately 0.05 e/f.u. doping value. Variation in the number of occupied states for the valence orbitals of (d) La, (e) W, and (f) N with charge doping. The circles denote electron doping, while the triangles indicate hole doping. Notably, the added charges occupy the valence orbitals of all three atoms.}
  \label{fig:DOS_all}
\end{figure*}

\begin{figure}[h]
  \centering
  \begin{subfigure}{0.495\textwidth}
    \centering
    \includegraphics[width=\linewidth]{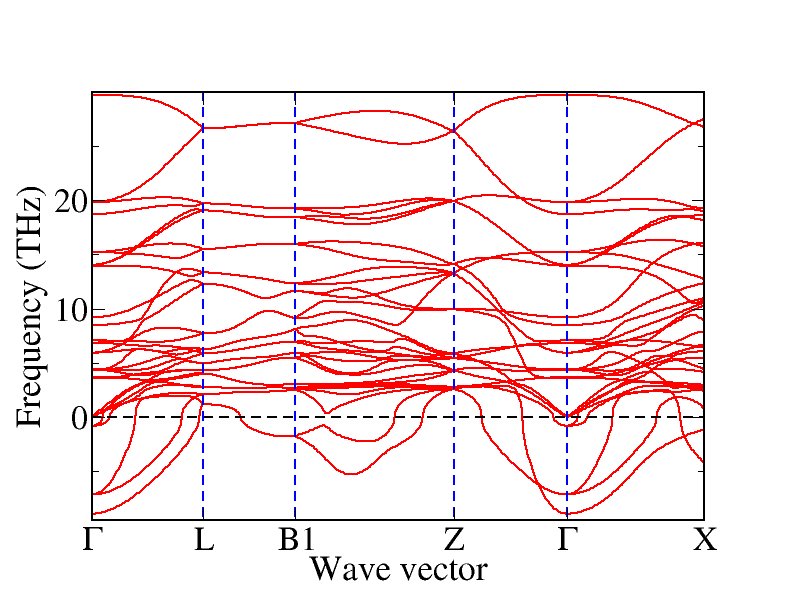}
    \caption{Undoped}
  \end{subfigure}
  \hfill
  \begin{subfigure}{0.495\textwidth}
    \centering
    \includegraphics[width=\linewidth]{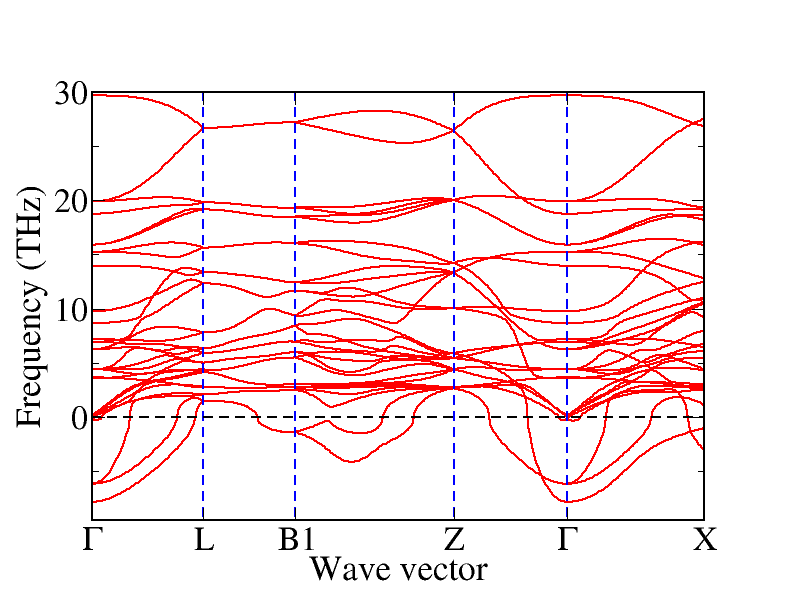} 
    \caption{0.10 e/f.u.}
  \end{subfigure}
  \hfill
  \begin{subfigure}{0.495\textwidth}
    \centering
    \includegraphics[width=\linewidth]{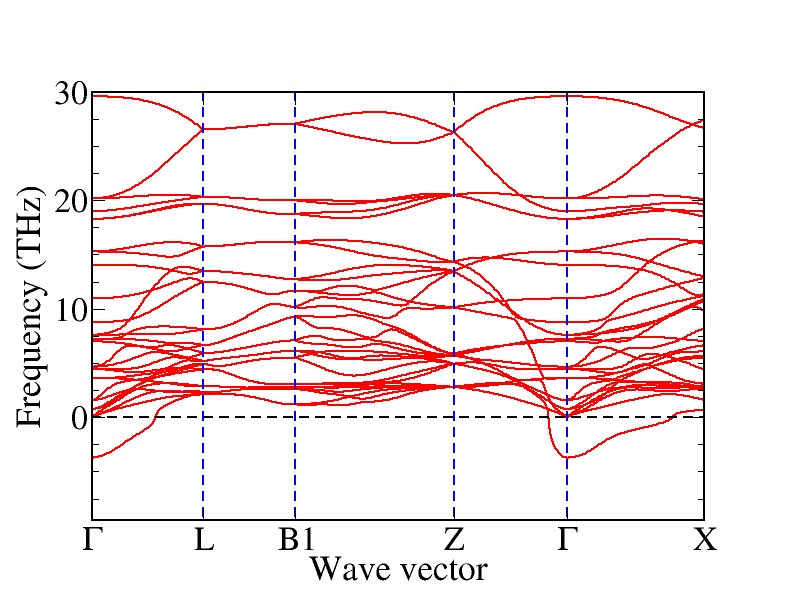} 
    \caption{0.15 e/f.u.}
  \end{subfigure}
  \hfill
    \begin{subfigure}{0.495\textwidth}
    \centering
    \includegraphics[width=\linewidth]{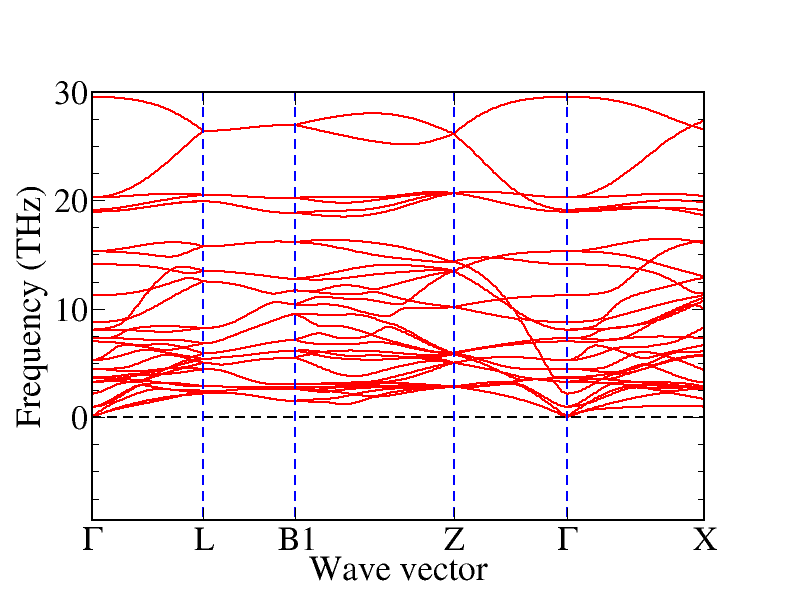} 
    \caption{0.20 e/f.u.}
    \label{fig:phonon_R3barc_0.30}
  \end{subfigure}
  \hfill
  \caption{Phonon dispersions for the R$\overline{3}$c structure with varying electron concentrations: (a) undoped, (b) 0.05 electrons/f.u., (c) 0.15 electrons/f.u., and (d) 0.20 electrons/f.u. At an electron concentration of 0.20 electrons/f.u., the absence of unstable (imaginary) modes in the phonon dispersion indicates a phase transition from a polar low symmetric structure to a non-polar high symmetric structure.}
  \label{fig:charge_phonon}
\end{figure}

We begin by evaluating the influence of charges on the frequency of the ferrodistortive unstable mode in the R$\overline{3}$c structure through the introduction of a background charge~\cite{wang2012ferroelectric,aschauer2014competition}, followed by assessing the change in polar distortion in R3c structure with charge doping. Notably, we find that the ferrodistortive modes remain unstable up to a substantial electron doping of 0.2 electrons/f.u. Here f.u. denotes formula unit. These unstable modes eventually become stabilized beyond a charge doping of 0.20 electrons/f.u. (Fig. \ref{fig:Back_ground_charge}a). Effect of charge doping on phonon dispersion along high symmetry points is shown in Fig. \ref{fig:charge_phonon}. We observe a suppression of polarization in R3c phase with the addition of charges, with a structural phase transition from a polar to a non-polar structure (low symmetry to high symmetry) occurring at approximately 0.21 electrons/f.u. (Fig. \ref{fig:Back_ground_charge}b). This value is consistent with the electron doping required to stabilize the ferrodistortive modes. Most remarkably, we discover that both the polar distortion and the unstable ferrodistortive mode are nearly unaffected by hole doping even up to a high hole concentration of 0.2 holes/f.u. Based on our observations from both electron and hole doping, we predict that substantial electron addition ($>0.2$ electrons/f.u.) will suppress ferroelectricity in LaWN\textsubscript{3} by mitigating the ferrodistortive distortions. 
In contrast, the introduction of holes does not significantly affect the polar distortions, as hole doping has a lesser impact on ferrodistortive modes compared to electron doping. This behavior in LaWN\textsubscript{3} parallels the response observed in certain oxide ferroelectric perovskites, such as BaTiO\textsubscript{3}, KNbO\textsubscript{3}, and BaMnO\textsubscript{3}~\cite{michel2021interplay,zhao2018meta}. However, there are some other ferroelectric materials that demonstrate distinct responses to charge carrier introduction. For instance, members of the A\textsubscript{n}B\textsubscript{n}O\textsubscript{3n+2} family, such as Sr\textsubscript{2}Nb\textsubscript{2}O\textsubscript{7} and La\textsubscript{2}Ti\textsubscript{2}O\textsubscript{7}, exhibit robust ferroelectric behavior that remains largely unaffected by either type of charge carrier addition~\cite{zhao2018meta}. Recently, studies on the geometric improper ferroelectric h-YMnO\textsubscript{3}, a member of the hexagonal manganite family, reveal an increase in polarization with electron doping and a suppression with hole doping~\cite{tosic2024interplay}. Therefore, it becomes important to understand the mechanism in nitride perovskites, as we elucidate next.

The change in polarization can be understood through electronic structure considerations. In Fig.~\ref{fig:DOS}, we plot the density of states (DOS) of the valence orbitals of the atoms (La, W, N) that significantly contribute to the states near the Fermi level. The states at the valence band maxima (VBM) are primarily composed of the $2p$ orbitals of N, whereas the conduction band minima (CBM) predominantly exhibit W $5d$ character. Notably, both the VBM and CBM have a significant density of states associated with the valence orbitals of other atoms.

Before proceeding further with the role of electronic structure, it is crucial to understand the arrangement of nitrogen around tungsten in the WN\textsubscript{6} octahedron for our subsequent analysis. As we previously discussed, W and N are distorted in opposite directions along the pseudocubic $\langle 111 \rangle$ direction (Fig.~\ref{fig:dist}). This results in W having equal bond lengths with three cis nitrogen atoms in the upper direction (grey-colored atoms) and equal bond lengths with the other three cis nitrogen atoms in the lower direction (yellow-colored atoms). These two pairs of cis nitrogen atoms have different bond lengths with W, where the bond strength of W-N (axial up) is greater than W-N (axial down), and the bond strength is inversely related to the bond length. For simplicity in our discussions, we will refer to the upper nitrogen atoms as axial up nitrogen and the lower ones as axial down nitrogen. We note that although there are two sets of nitrogen atoms within a particular unit cell, they are all equivalent because the axial up nitrogen atoms transform to axial down nitrogen atoms for the next tungsten in the neighboring unit cell.

Returning to our discussion, Fig.~\ref{fig:DOS_h} and Fig.~\ref{fig:DOS_e} illustrate the effect of hole and electron doping on the partial density of states (PDOS). We find that with the addition of holes, the density of states are not significantly affected, apart from the shift of the Fermi level into the valence band manifold. However, with the addition of electrons, instead of a rigid shift of the Fermi level, the band gap starts to vanish by ``filling up" with states. Nonetheless, in both cases, our system transitions to a metallic state, as expected.

\begin{figure}
  \centering

  \begin{subfigure}{0.48\textwidth}
    \centering
    \includegraphics[width=\linewidth]{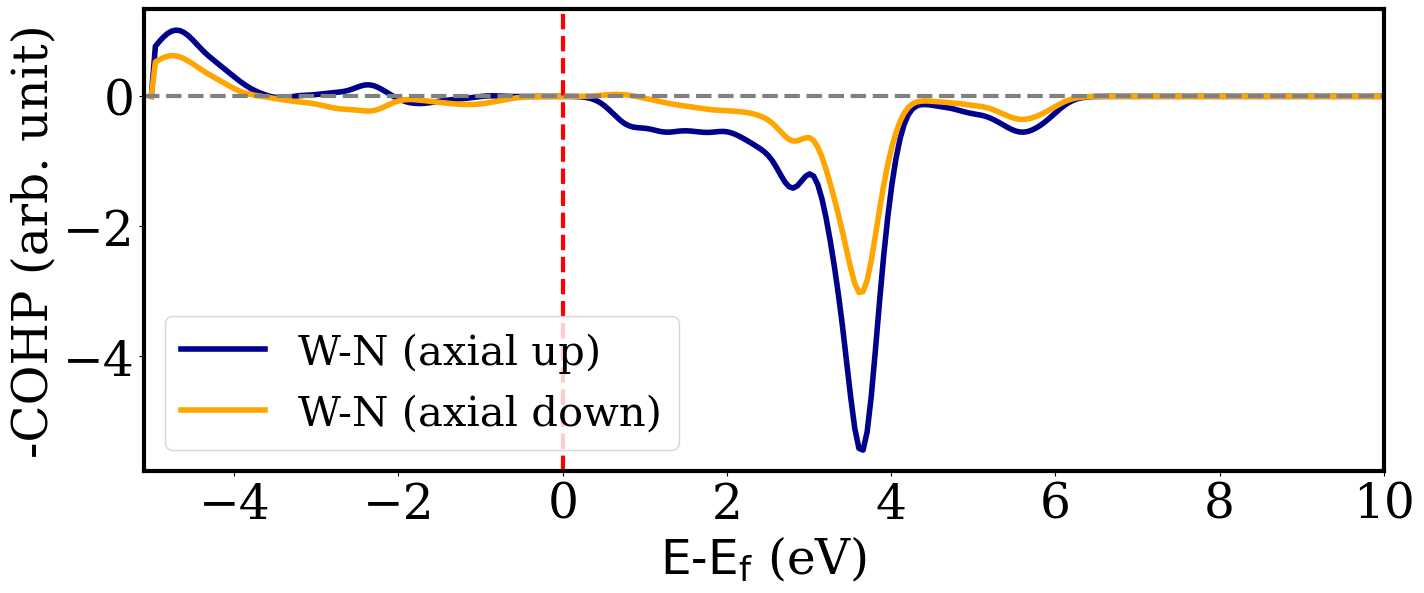}
    \caption{}
    \label{fig:COHP}
  \end{subfigure}
  \hfill
  \begin{subfigure}{0.48\textwidth}
    \centering
    \includegraphics[width=\linewidth]{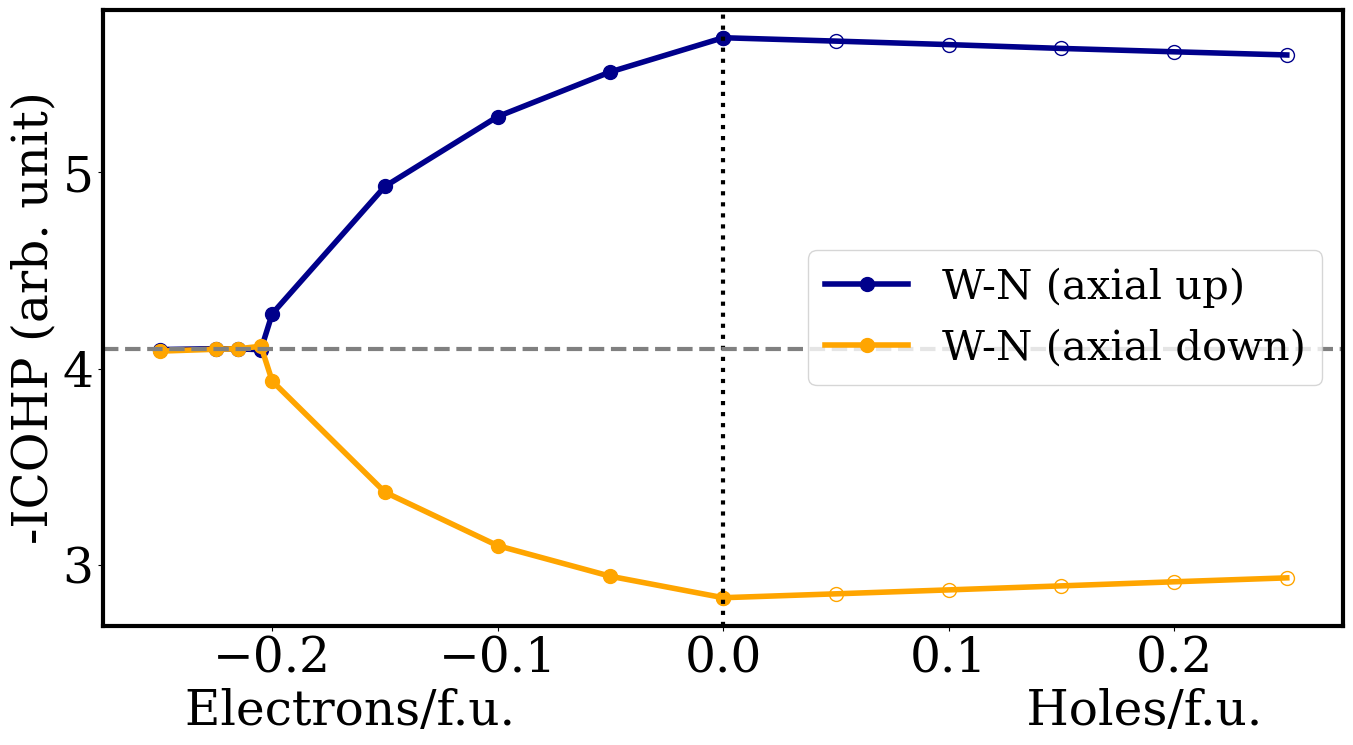}
    \caption{}
    \label{fig:ICOHP}
  \end{subfigure}
  \hfill
  \caption{a) -COHP between W and axial up N (blue) and W and axial down N (yellow). Negative values indicate antibonding states, whereas positive values represent bonding states. b) -ICOHP illustrating the bond strength between W-N (axial up) and W-N (axial down) in blue and yellow color respectively. The bond strength of W-N (axial up) equals that of W-N (axial down) at 0.21 e/f.u., while the addition of holes has a negligible impact.}

\end{figure}

In Figs.~\ref{fig:occ_La}, \ref{fig:occ_W}, and \ref{fig:occ_N}, we plot the number of occupied states with the doping concentration. Unlike in the ferroelectric BaTiO\textsubscript{3}, where added electrons occupy the $3d$ orbitals of the B-site Ti atom and added holes occupy the $2p$ orbitals of O~\cite{michel2021interplay}, in this case, both added electrons and holes occupy the orbitals of all three atoms (La, W, N) due to their significant contribution to the states near the Fermi level.

To determine how the bonding between W and N evolves with doping, we evaluate the COHP for the undoped compound (as shown in Fig.~\ref{fig:COHP}), as well as the ICOHP up to the Fermi level as a function of doping concentration (shown in Fig.~\ref{fig:ICOHP}). From the COHP, we find that the empty electronic states above the Fermi level exhibit significant W-N (axial up) anti-bonding character, as shown in Fig.~\ref{fig:COHP}. This suggests that the addition of electrons decreases the bonding between W and the axial up N atoms, and also reduces the off-centering of W. Conversely, the filled electronic states below the Fermi level show a negligible W-N (axial down) antibonding character. Consequently, the addition of holes tends to increase the bonding between W and the axial down N atoms, which eventually decreases the off-centering of W, albeit by a negligible amount. This mechanism is further confirmed by examining the -ICOHP for W-N (axial up) and W-N (axial down), which become equal to each other at nearly 0.21 electrons/f.u. (see Fig.~\ref{fig:ICOHP}). This is consistent with the concentration of electrons required for the phase transition from a low-symmetry structure to a high-symmetry structure, as shown in Fig.~\ref{fig:Back_ground_charge}. On the other hand, the introduction of holes has a negligible impact on the -ICOHP, as well as the polar distortion in the R3c structure. Furthermore, the robustness of polar distortion under hole doping can be understood by considering the high electron density in the $2p$ orbitals of nitrogen. After the addition of holes (removal of electrons), nitrogen still retains a sufficiently high electron density in its $2p$ orbitals to overlap effectively with the $5d$ orbitals of W. Additionally, a significant portion of the added holes occupy the $5d$ orbitals of W, which further reduces the repulsion between the electron clouds of W and N.

To summarize, we have demonstrated that the addition of electrons suppresses the polar distortions in the R3c structure of LaWN\textsubscript{3}, with approximately 0.21 electrons/f.u. needed to reach a non-polar state. Similarly, the addition of holes also suppresses polarization, but to a negligible extent. We predict that a phase transition from the R3c structure will not occur at any experimentally-reachable hole concentration. We have delineated the underlying mechanism of this robustness under charge doping, and propose that perovskite nitride LaWN\textsubscript{3} is a promising polar metal candidate. Next, we move on to the study of impurity atoms to further understand the fate of polar distortions under doping.

\begin{figure*}[t]
  \centering
  \includegraphics[width=0.9\linewidth]{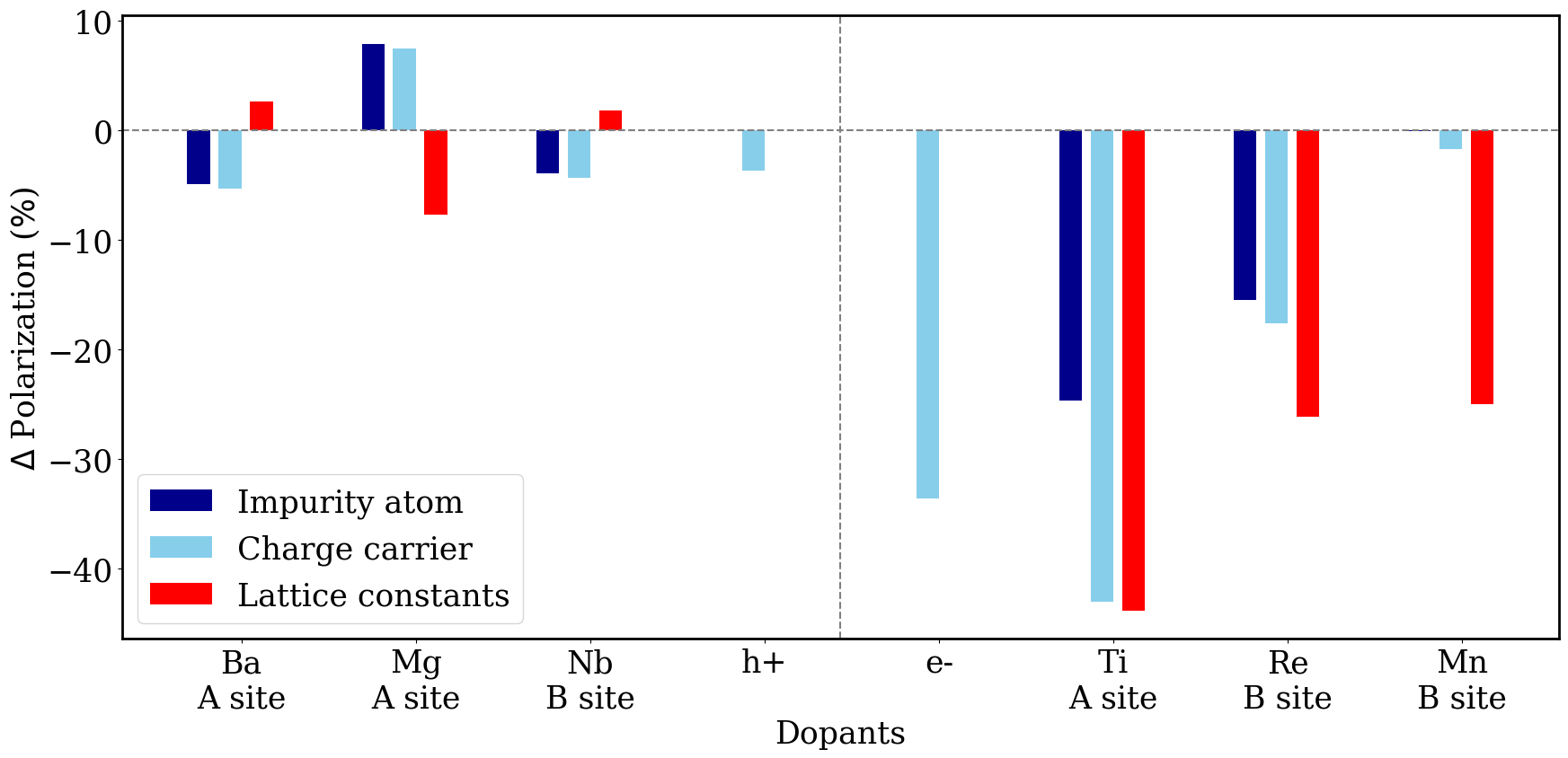} 
  \caption{Percentage change in polarization of the R3c structure of LaWN\textsubscript{3} with the introduction of explicit dopants in the $2 \times 2 \times 1$ supercell, contributing 0.125 charge per formula unit to the system. Negative values indicate a reduction in polarization, while positive values indicate an enhancement. The three factors influencing polarization are shown: (i) Impurity atom contribution (in dark blue), (ii) Charge carrier contribution (in sky blue), and (iii) Lattice constant contribution (in red). Note that Ba, Mg, and Nb add holes to the system, whereas Ti, Re, and Mn add electrons. The results from background charges is also shown indicated by h+ and e-.} \label{fig:Impurity_atoms}
\end{figure*}

\subsection{Impurity atoms in supercells}
\label{sec:Impurity atoms in supercells}

Following our study of background charge doping, we next introduce impurity atoms into the R3c structure of LaWN$_3$ and examine the resultant modifications in polar distortions and polarization. Specifically, we substitute one La or W atom by different impurity atoms, one at a time, in a 2 $\times$ 2 $\times$ 1 supercell, equating to the replacement of 1 out of 40 atoms per instance. For electron doping, La at the A-site is replaced with Ti, while W at the B-site is substituted with Re and Mn. For hole doping, La is replaced with Ba and Mg, and W is substituted with Nb. Each of these impurity atoms contributes an additional charge (either hole or electron) to the system, corresponding to 0.125 charge/f.u. These doped systems are anticipated to exhibit intricate alterations relative to background charge calculations due to the disparity in atomic size between the impurity and the substituted atom, as well as the concomitant volumetric changes in the unit cell. To disentangle the contributions of various factors to the polarization change, we conduct three separate sets of calculations, focusing on the contributions from the impurity atoms, charge carriers, and lattice constants, as elucidated in Refs.~\cite{michel2021interplay,tosic2024interplay}. We employ Shannon ionic radii~\cite{shannon1976acta} to estimate the sizes of the impurity atoms in their respective oxidation states and coordination environments. We next examine each of the above-mentioned contributions.

\subsubsection{Impurity atom contribution}

In the first set of calculations, we aim to discern the influence of atomic size on the polarization of X-LaWN$_3$, where X represents the impurity atom. We replace La or W with a dopant and introduce compensating background charge to maintain system neutrality, thus isolating the effect of the dopant's size from its charge contribution. The lattice constants are held constant at the values of undoped LaWN$_3$, and internal atomic coordinates are relaxed.

Our results are depicted by the dark blue-colored bars in Fig.~\ref{fig:Impurity_atoms}. We find that for the dopants introducing holes into the system, an expected trend emerges -- as the size of the dopant increases relative to the substituted atom, polarization decreases. A smaller atom at the A-site provides more space for W to off-center, as observed in Mg-LaWN$_3$, while a larger atom, such as in Ba-LaWN$_3$, restricts this off-centering. Additionally, Nb$^{5+}$, due to its larger size compared to W$^{6+}$, becomes less off-centered and restricts the off-centering of adjacent W atoms, leading to a reduction in polarization. Conversely, this size dependency is not observed for dopants that introduce electrons into the system, as all these atoms are smaller than the substituted atoms. This anomaly arises from significant alterations in the contribution of orbitals near the Fermi level. While hole-donating dopants do not substantially affect the PDOS, electron-donating dopants exhibit a significant presence at the Fermi level, exceeding that of the atoms they replace. This increased orbital contribution, coupled with the smaller size of the dopants, results in the observed deviations from the expected trend. The graphs of the PDOS for X-LaWN\textsubscript{3} are presented in Fig~\ref{fig:Impurity_dos}.

\subsubsection{Charge carrier contribution}

In the second set of calculations, we examine how the size of the dopant and its associated charge carriers jointly influence the polarization. This method can be compared with the background charge method (with an added charge of 0.125 per f.u.), where additional charge was introduced without any dopant. The differences in polarization change observed in this method, compared to the background charge approach can be attributed to the extra features of the impurity atom, such as size difference and valence orbital contributions to states near the Fermi level. For these calculations, we do not add compensating background charge, although the lattice constants are kept the same as the undoped structure. Our results are shown by the sky blue-colored bars in Fig.~\ref{fig:Impurity_atoms}.

\begin{figure}[h]
  \centering
  \begin{subfigure}{0.32\textwidth}
    \centering
    \includegraphics[width=\linewidth]{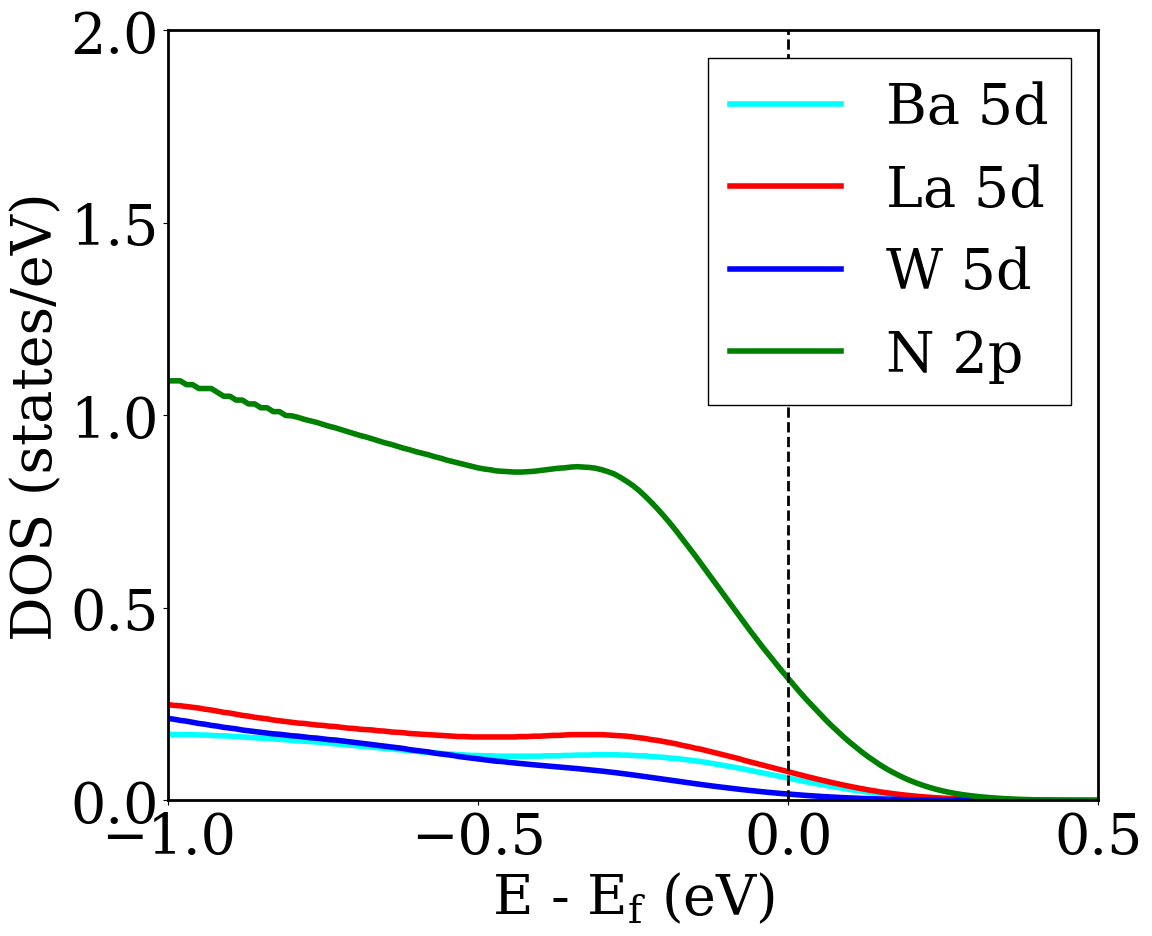}
    \caption{Ba-LaWN\textsubscript{3}}
    \label{fig:La_Ba}
  \end{subfigure}
  \hfill
  \begin{subfigure}{0.32\textwidth}
    \centering
    \includegraphics[width=\linewidth]{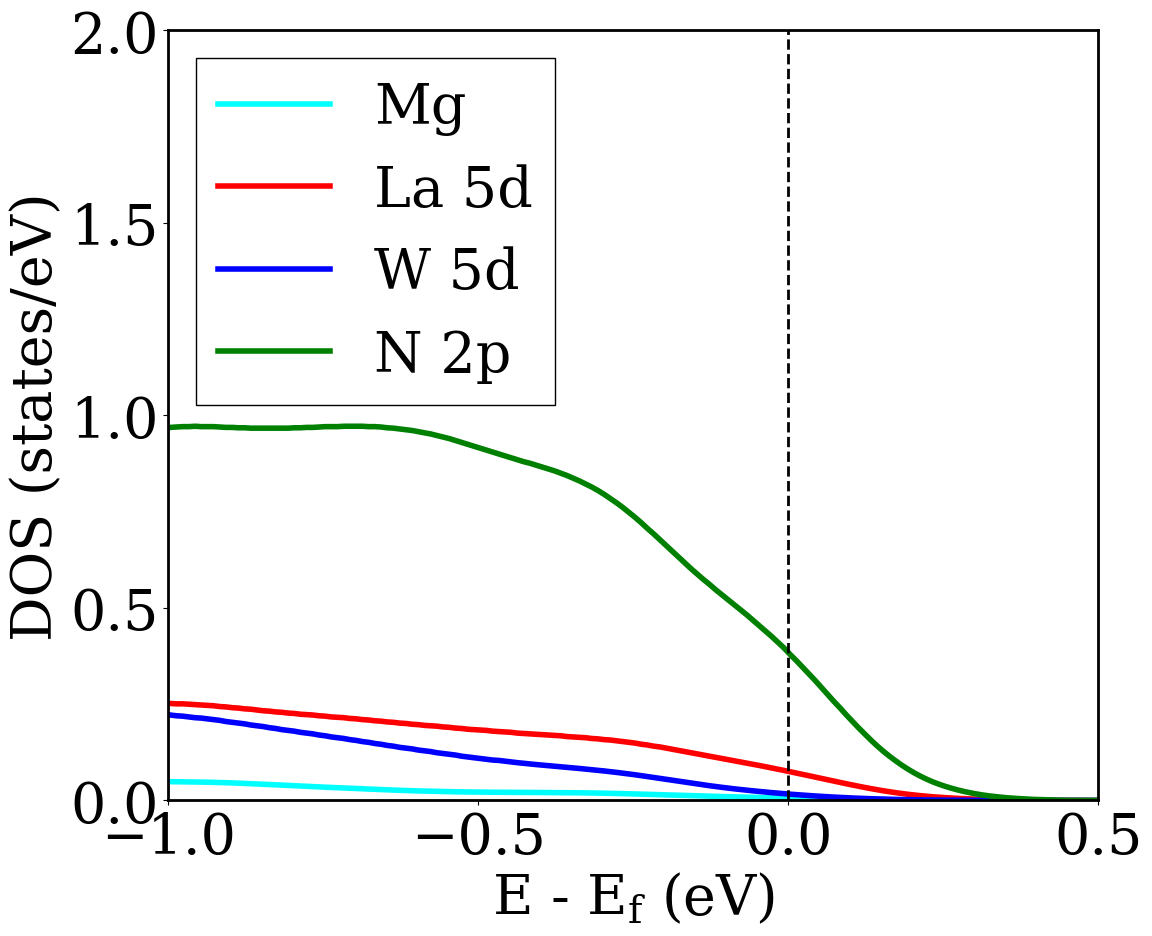} 
    \caption{Mg-LaWN\textsubscript{3}}
    \label{fig:La_Mg}
  \end{subfigure}
  \hfill
  \begin{subfigure}{0.32\textwidth}
    \centering
    \includegraphics[width=\linewidth]{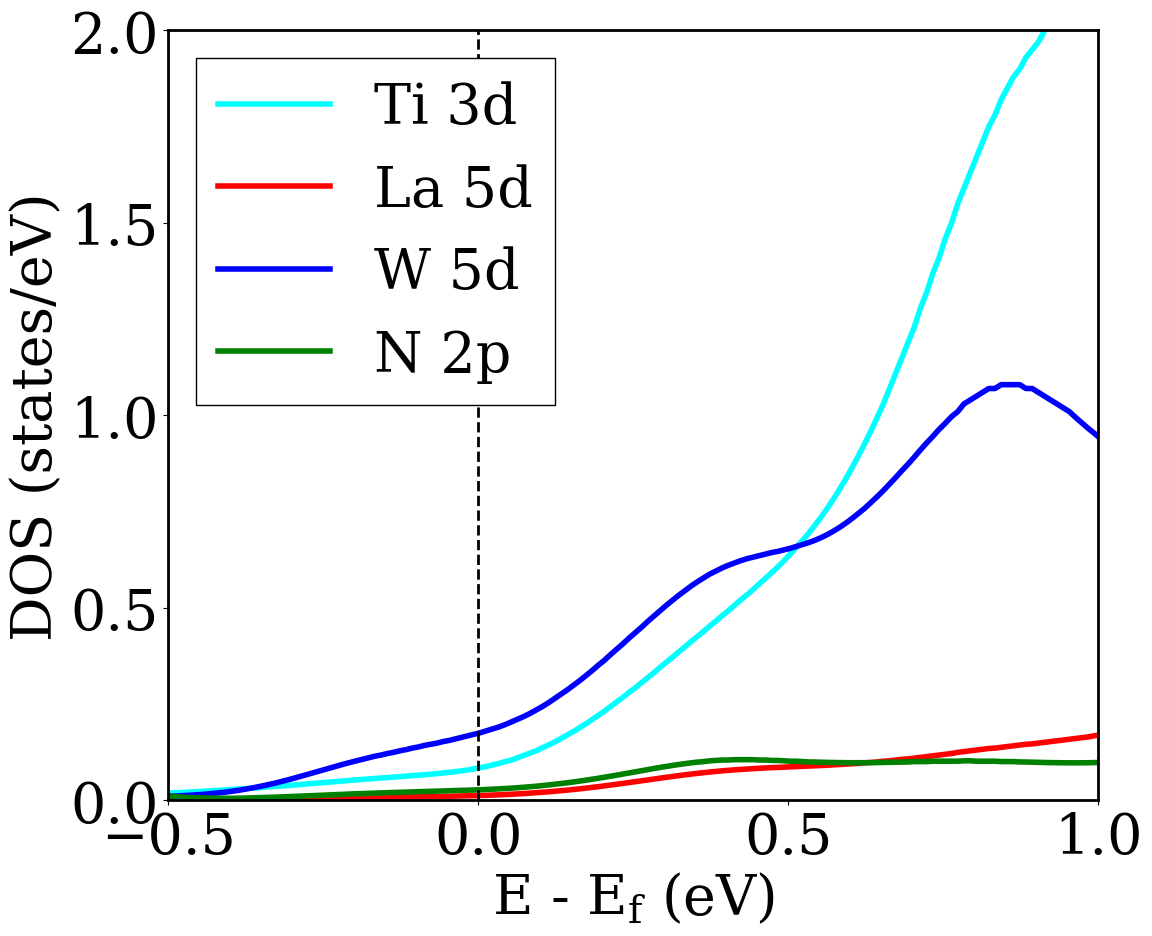} 
    \caption{Ti-LaWN\textsubscript{3}}
    \label{fig:La_Ti}
  \end{subfigure}
  \hfill
    \begin{subfigure}{0.32\textwidth}
    \centering
    \includegraphics[width=\linewidth]{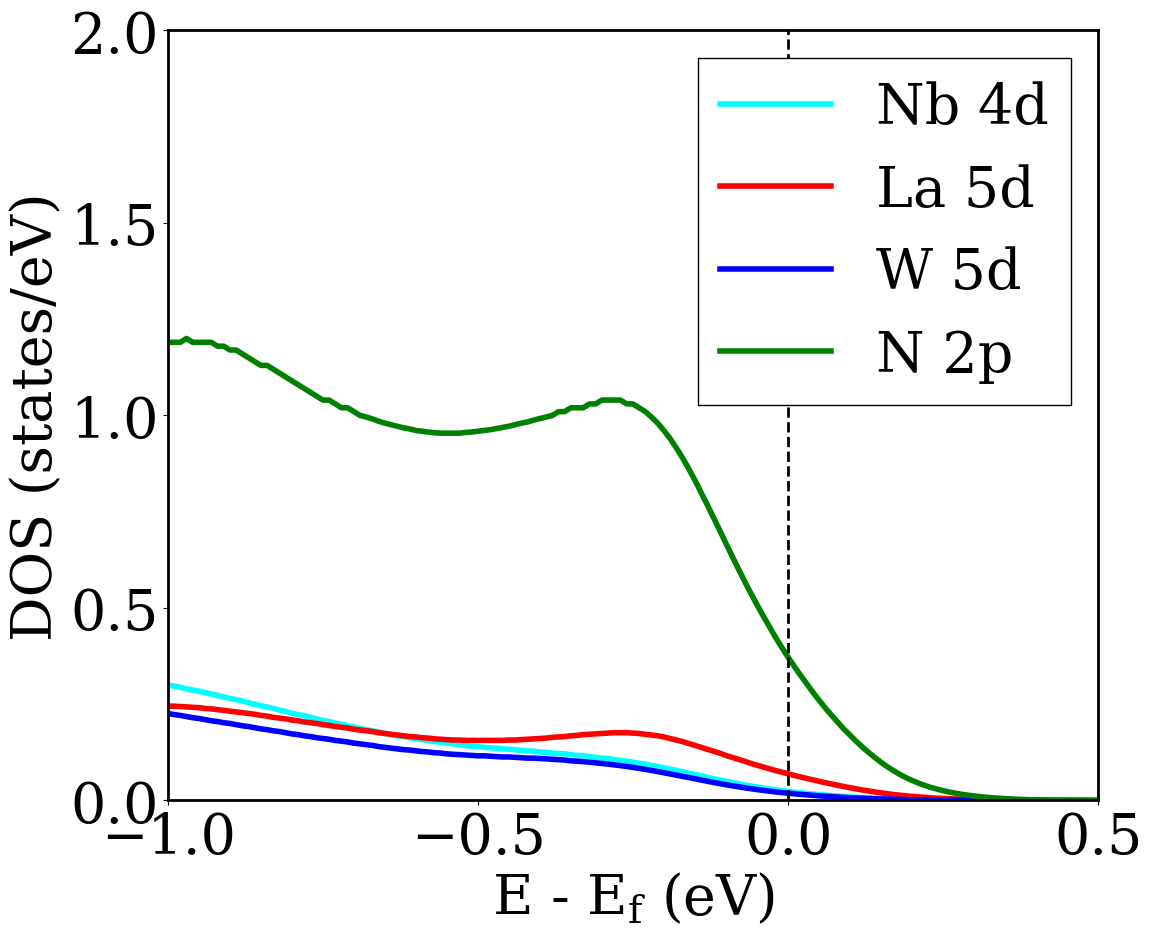} 
    \caption{Nb-LaWN\textsubscript{3}}
    \label{fig:W_Nb}
  \end{subfigure}
  \hfill
  \begin{subfigure}{0.32\textwidth}
    \centering
    \includegraphics[width=\linewidth]{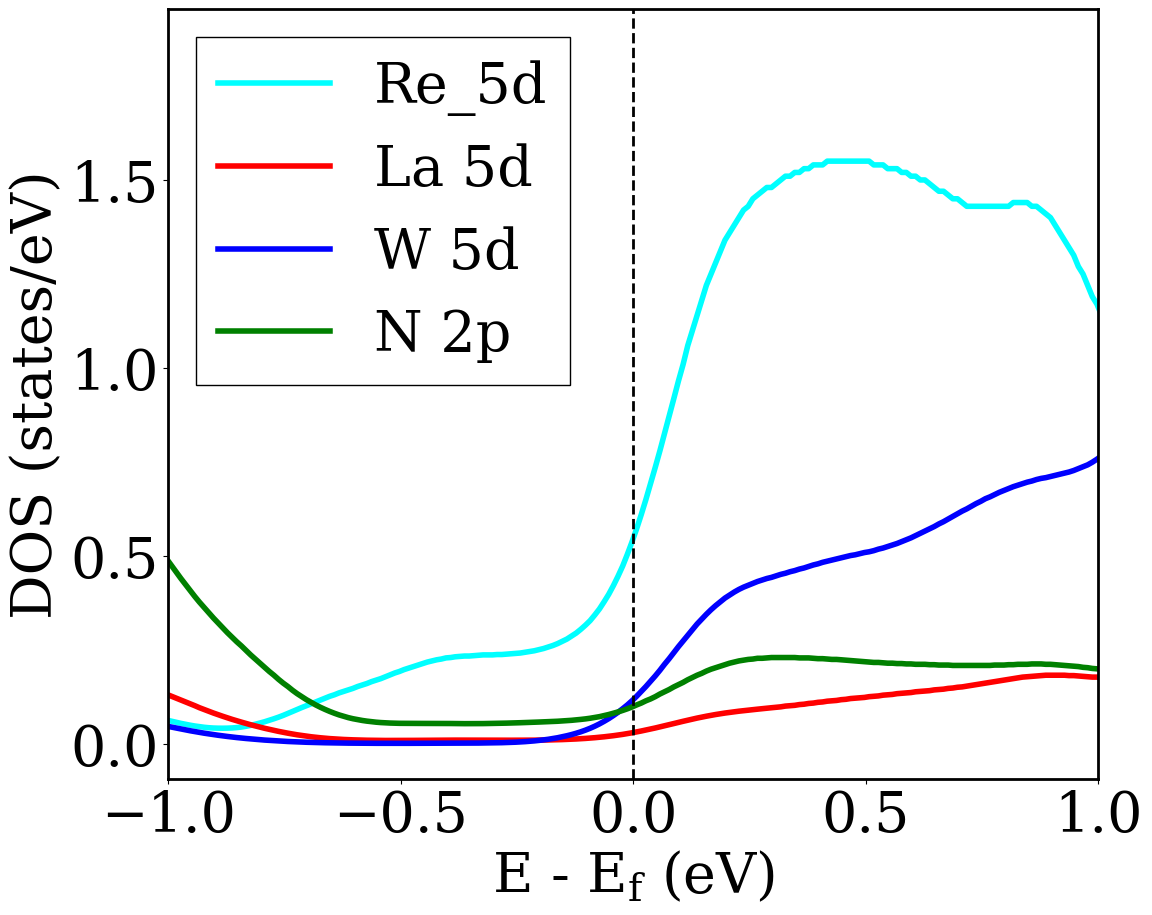} 
    \caption{Re-LaWN\textsubscript{3}}
    \label{fig:W_Re}
  \end{subfigure}
  \hfill
    \begin{subfigure}{0.32\textwidth}
    \centering
    \includegraphics[width=\linewidth]{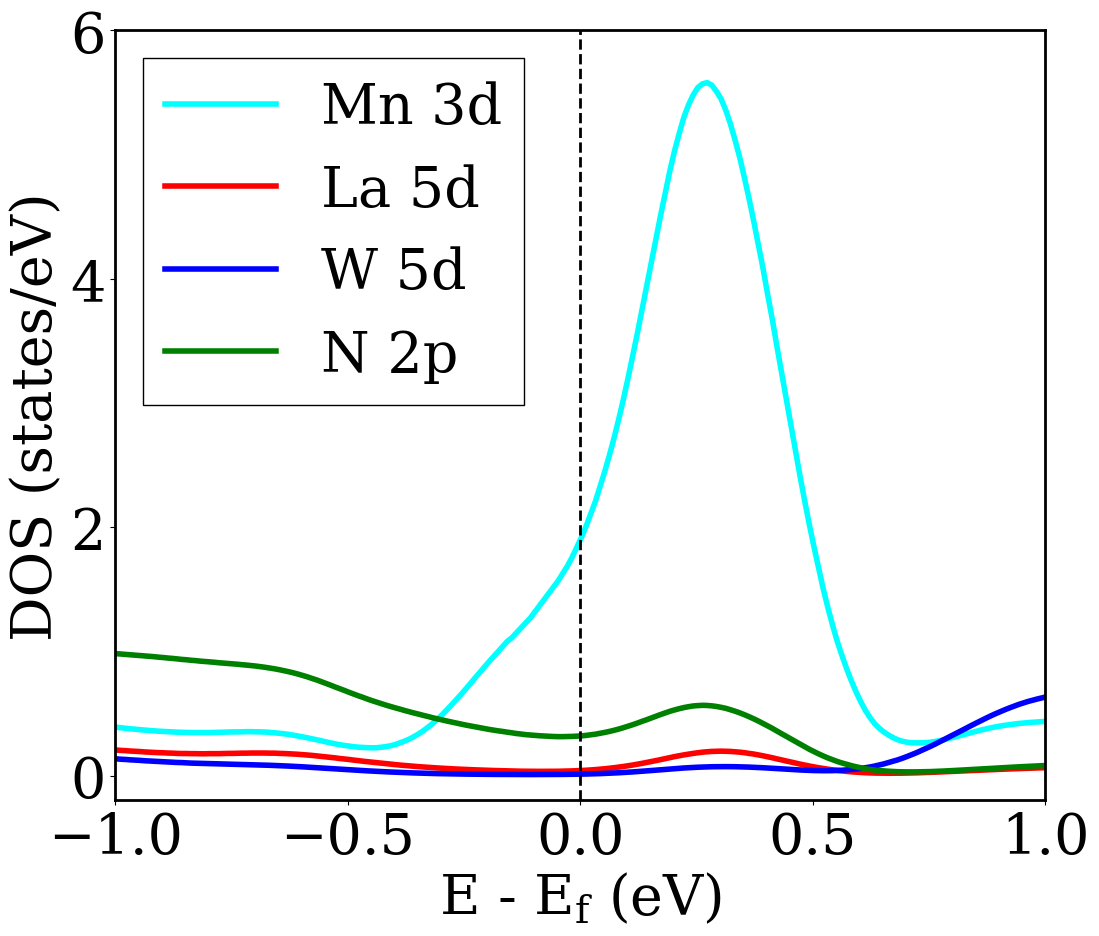} 
    \caption{Mn-LaWN\textsubscript{3}}
    \label{fig:W_Mn}
  \end{subfigure}
  \hfill
  \caption{The density of states (DOS) for the valence orbitals of X-LaWN\textsubscript{3} are presented for various dopants. (a) Ba and (d) Nb dopants exhibit DOS profiles that closely resemble those of the substituted atoms, indicating that they do not significantly alter the electronic structure of the host material. In contrast, (b) Mg shows minimal contribution to the states near the Fermi level. (c) Ti introduces a slightly increased contribution near the Fermi level compared to the original atoms. Notably, (e) Re and (f) Mn dopants display substantially higher contributions to the states near the Fermi level, indicating significant electronic interactions and pronounced modifications to the electronic structure.}
  \label{fig:Impurity_dos}
\end{figure}

As discussed for acceptor dopants (Ba$^{2+}$, Mg$^{2+}$, Nb$^{5+}$), there is no significant change in the partial density of states (Fig.~\ref{fig:La_Ba},~\ref{fig:La_Mg},~\ref{fig:W_Nb}). However, there is a slight increase in hole accumulation in the $2p$ orbitals of N, which leads to a decrease in the off-centering of the B-site atom due to its weaker bonding with N. This results in a slight decrease in polarization in Ba-LaWN$_3$ and Nb-LaWN$_3$, compared to the background charge method. The increase in polarization in Mg-LaWN$_3$ can be explained by the dominance of the impurity atom's contribution to the polarization change over the charge carrier effect for acceptor dopants.

For donor dopants, replacing La$^{3+}$ with Ti$^{4+}$ results in a greater reduction in polarization than observed with the background charge method. This can be understood as the contribution of the impurity atom adds up with the contribution from added electrons -- this leads to a greater reduction. However, the polarization is less affected in Re-LaWN$_3$ and even less so in Mn-LaWN$_3$. This is due to the dominant contribution of the $5d$ orbitals of Re and the $3d$ orbitals of Mn near the Fermi level (shown in Fig.~\ref{fig:W_Re} and Fig.~\ref{fig:W_Mn}). The added electrons primarily accumulate in these orbitals, leading to fewer electrons in the $5d$ orbitals of W, thus minimally affecting the W-N bonding. This results in a lesser change in polarization compared to the background charge method. The difference in polarization change between Re-LaWN$_3$ and Mn-LaWN$_3$ is related to the quantitative differences in the density of their respective valence orbitals near the Fermi level.

\subsubsection{Lattice constant contribution}

Finally, in the third set of calculations, we perform a complete relaxation of both the internal coordinates and the lattice constants after introducing the dopant with charge carriers. This set of calculations represents the most realistic scenario, as it accounts for the combined contribution of all the factors. The results are depicted by the red-colored bars in Fig.~\ref{fig:Impurity_atoms}. We find an increase in polarization with the addition of Ba$^{2+}$ dopant at the A site and Nb$^{5+}$ dopant at the B site. This is attributed to their larger size compared to the atoms they substitute (La and W, respectively), which increases the unit cell volume, thereby providing more space for the B-site atom to exhibit greater off-centering. In contrast, for the other dopants, the smaller size compared to the atoms they replace leads to a decrease in volume, resulting in less space for the off-centering of the B-site atom. This explains the decrease in polarization observed for these dopants. The magnitude of the polarization change from this contribution can be understood by considering all factors, such as the change in volume related to the size of the dopant and how charge carriers affect the bonding between W and N.

Let us summarize our key findings regarding the factors influencing the polarization upon doping. We saw that for dopants that do not significantly alter the contributions of other orbitals near the Fermi level, the polarization increases if the dopant is smaller than the substituted atoms and decreases if the dopant is larger. However, dopants such as Ti, Re, and Mn deviate from this size-dependent trend due to their significant contributions to states near the Fermi level. When examining the charge carrier contribution, an interesting behavior is observed in Re and Mn doped LaWN\textsubscript{3}. Due to their dominant contribution in the states near the Fermi level, the bonding between W and N remains relatively unaffected, resulting in a smaller decrease in polarization compared to the results obtained using the background charge method. Finally, by analyzing the lattice constant contribution, we found that the unit cell volume, which is influenced by the size of the dopant, plays a crucial role in the observed changes in polarization.

\section{Summary and conclusions}
\label{sec:Conclusions}

To summarize, we have demonstrated that recently-synthesized perovskite nitride LaWN\textsubscript{3} can sustain both polarity and metallicity. Using the background charge method, we found that the polarization of LaWN\textsubscript{3} is suppressed with the addition of approximately 0.21 electrons per formula unit. The addition of holes also suppresses polarization but to a very small extent, making the polar structure remarkably robust to hole doping. We presented the detailed mechanism behind this occurrence -- added electrons decrease the bonding strength between W and axial up N because electrons occupy states with W-N (axial up) antibonding character. In contrast, when adding holes, electrons were removed from states that do not significantly affect the W-N (axial up) bonding, resulting in a negligible impact on this bonding. Since the introduction of impurity atoms affects polarization through various factors, we isolated three contributions that influence the change in polarization -- the chemistry and the size of the impurity atom, electronic effects shown by charge carriers, and changes in lattice constants. This gives a complete picture of the interplay between polarity and metallicity in this prototypical perovskite nitride. We conclude that X-LaWN\textsubscript{3} can be a promising polar metal, especially if X is an acceptor dopant, as the addition of holes does not significantly decrease the polar distortion. We hope our findings will be useful for the design and synthesis of polar metallic nitride perovskites.

\section*{Acknowledgments}

HSD acknowledges Indian Institute of Science for a fellowship. AN acknowledges support from ANRF (project number CRG/2023/000114).

\bibliography{ref.bib}

\pagebreak
\widetext
\newpage
\begin{center}
	\textbf{\Large Appendix}
\end{center}

\section{Phonon dispersions with electron doping}

\newpage
\section{Density of states of X-L\MakeLowercase{a}WN\textsubscript{3}}

\end{document}